\documentclass[twoside]{IEEEtran}

\usepackage{verbatim}
\usepackage{varioref}
\usepackage{subfigure}

\usepackage{psfrag}
\usepackage{epsf}
\usepackage{graphics}
\usepackage{graphicx}
\usepackage{amsbsy}
\usepackage{amssymb}
\usepackage{amscd}
\usepackage{amsmath}
\usepackage{enumerate}
\usepackage{algorithm2e}
\usepackage{color}

\usepackage{threeparttable}
\usepackage{algorithm2e}

\usepackage{cite}

\newtheorem{lemma}{Lemma}

\newtheorem{remark}{Remark}

\newtheorem{definition}{Definition}

\def\mD{{{\mathcal{D}}}}
\def\mE{{{\mathcal{E}}}}
\def\mR{{{\mathcal{R}}}}
\def\mS{{{\mathcal{S}}}}

\def\mU{{{\mathcal{U}}}}
\def\mV{{{\mathcal{V}}}}
\def\mX{{{\mathcal{X}}}}
\def\mY{{{\mathcal{Y}}}}
\def\mZ{{{\mathcal{Z}}}}

\def\mbC{{\boldsymbol{\mathcal{C}}}}

\def\mbN{{\boldsymbol{\mathcal{N}}}}
\def\mbW{{\boldsymbol{\mathcal{W}}}}

\def\bx{{\boldsymbol{x}}}
\def\bX{{\boldsymbol{X}}}

\def\bz{{\boldsymbol{z}}}

\def\bZ{{\boldsymbol{Z}}}

\def\bR{{\boldsymbol{R}}}

\def\by{{\boldsymbol{y}}}
\def\bY{{\boldsymbol{Y}}}

\def\bH{{\boldsymbol{H}}}
\def\bh{{\boldsymbol{h}}}

\def\bw{{\boldsymbol{w}}}

\title{Scalable Capacity Bounding Models for Wireless Networks}

\author{Jinfeng Du, \emph{Member, IEEE}, Muriel M{\'e}dard, \emph{Fellow,
IEEE}, Ming Xiao,
\emph{Senior Member, IEEE},\\
 and Mikael Skoglund, \emph{Senior Member, IEEE}
\thanks{This work was presented in part at the IEEE International Symposium on Information Theory in July 2013 and in July 2014. This work was funded in part by the Swedish Research Council (VR) and by the Air Force Office of Scientific Research (AFOSR) under award No. FA9550-09-1-0196 and FA9550-13-1-0023.
 }
\thanks{Jinfeng Du is with Research Lab of Electronics, Massachusetts Institute of Technology,
Cambridge, MA, USA (Email: jinfeng@mit.edu). He was with School of Electrical Engineering, Royal Institute of Technology,
Stockholm, Sweden.}
\thanks{Muriel M\'edard is with Research Lab of Electronics, Massachusetts Institute of Technology,
Cambridge, MA, USA (Email: medard@mit.edu).}
\thanks{Ming Xiao and Mikael Skoglund are with
 School of Electrical Engineering and the ACCESS Linnaeus Center, Royal Institute of Technology,
Stockholm, Sweden (Email: \{mingx, skoglund\}@kth.se).}
\thanks{Copyright (c) 2014 IEEE. Personal use of this material is permitted.  However, permission to use this material for any other purposes must be obtained from the IEEE by sending a request to pubs-permissions@ieee.org.}
}

\begin{document}

\maketitle

\begin{abstract}

 The framework of network equivalence theory developed by  Koetter et al.  introduces a notion of channel emulation to construct noiseless networks as upper (resp. lower) bounding models, which
 can be used to calculate the outer (resp. inner) bounds for the capacity region of the original noisy network.
 Based on the network equivalence framework, this paper presents scalable upper and lower  bounding models for wireless networks with potentially many nodes. A channel  decoupling method is proposed to decompose wireless networks into decoupled multiple-access channels (MACs) and broadcast channels (BCs). The
 upper   bounding model, consisting of only point-to-point  bit pipes, is constructed  by firstly extending the ``one-shot'' upper  bounding models developed by Calmon et al.  and then integrating them with network equivalence tools. The  lower bounding model, consisting of both  point-to-point and
 point-to-points bit pipes, is constructed based on a two-step
 update of the lower bounding models to incorporate the broadcast nature
 of wireless transmission. The main advantages of the proposed
 methods are  their simplicity and the fact that they can be
 extended easily to large  networks with a complexity that grows linearly with the number of nodes.
 It is demonstrated that the  resulting upper and lower bounds can approach the capacity in some setups.
\end{abstract}

\begin{IEEEkeywords}
 capacity, channel decoupling, channel emulation, equivalence,  wireless networks
\end{IEEEkeywords}

\section{Introduction}\label{sec:Eqv:intro}

A theory of network equivalence has been established in~\cite{NetEquiv,NetEquiv2} by
 Koetter et al. to characterize the capacity of a (large) memoryless noisy network. The original noisy network is
first decomposed into many independent single-hop noisy channels,
each of which is then replaced by the corresponding upper (resp. lower) bounding  model consisting of
only noiseless bit pipes. The capacity region of the resulting noiseless network serves as an outer (resp. inner) bound for the capacity region of the original noisy network, whose capacity is otherwise difficult to
characterize. A noisy channel and a noiseless
bit pipe are said to be equivalent if the capacity region of any arbitrary
network that contains the noisy channel remains unchanged after
replacing the noisy channel by its noiseless counterpart. The
equivalence between a point-to-point noisy channel and a
point-to-point noiseless bit pipe has been  established in~\cite{NetEquiv}
as long as the rate of the latter equals the capacity of the
former. For independent single-hop multi-terminal channels, such
as the multiple-access channel (MAC), the broadcast channel (BC),
and the interference channel (IC),  operational frameworks for
constructing upper and lower bounding models have been proposed
in~\cite{NetEquiv2}. The constructive proofs presented in~\cite{NetEquiv,NetEquiv2} are based on a notion of channel emulation over a stacked replicas of the channel, where the lower bounding models are established
based on  channel coding arguments and the upper bounding models are constructed based on
lossy source coding arguments.

The bounding accuracy, in terms of both multiplicative and
additive gaps between capacity upper and lower bounds, has been
outlined in~\cite{NetEquiv2} for general noisy networks. Explicit
upper and lower bounding models for MAC/BC/IC with two sources
and/or two destinations have been constructed
in~\cite{NetEquiv2,Effros1,Effros2}. For networks consisting of
only point-to-point channels, MACs with two transmitters, and BCs
with two receivers,  the additive gap for Gaussian networks and
the  multiplicative gap for binary networks have been specified
in~\cite{Effros1}. The bounds obtained from network equivalence
tools~\cite{NetEquiv2,Effros1,Effros2} can be tight in some
setups, as shown in~\cite{Effros2} for a multiple unicast network
 consisting of noisy two-user BCs, and
in~\cite{Fawaz11} for a frequency-division AWGN relay network in
the wideband regime when the BC is physically degraded or when the
source treats the stochastically degraded BC as physically
degraded. A class of ``one-shot'' upper bounding models proposed
in~\cite{FlavioITW} by Calmon et al. introduces an auxiliary node for each BC/MAC
 to facilitate separate characterization of the sum rate and the individual rates.
The rate of a bit pipe from/to the auxiliary node can be characterized either by channel emulation
over infinite number of channel uses as in~\cite{NetEquiv,
NetEquiv2}, or by emulating the transmission over each channel use (hence named
``one-shot'').

Although it has been established in \cite{Shannon48} that separate source and channel coding incurs no capacity loss in transmitting a memoryless source over a  memoryless point-to-point channel, similar separation results on memoryless networks are not known until recently. The optimality of channel and network coding separation for general communication demands over networks consisting of independent memoryless point-to-point channels is established in~\cite{NetEquiv} based on the network equivalence theory, where the networks may be cyclic or acyclic and alphabets can be discrete or continuous. Feedback and cooperation among nodes are also accommodated, but the sources have to be independent.  For single-source multicast connections over networks consisting of point-to-point discrete memoryless channels (DMCs) with finite alphabets, separation of channel and network coding has been established in~\cite{Borade02} for acyclic synchronized networks and in~\cite{Song06} for asynchronized networks with and without cycles. While both~\cite{Borade02} and~\cite{Song06} rely on the max-flow min-cut theorem \cite{CoverThomas}  to show the converse (cf. upper bounding model), similar results  on channel-network separation have been established in~\cite{Hassibi07} by using normalized entropy vectors.
For point-to-point channels, the same upper bounding models established in~\cite{NetEquiv} have also been developed in~\cite{BSST02} for DMCs with finite-alphabet, and in~\cite{PCuff10} under the notion of strong coordination, where
 total variation (i.e., an additive gap) is used  to measure the difference between the desired
joint distribution and the empirical joint distribution of a pair of sequences (or a pair of symbols as in \emph{empirical coordination}).
 The concept of channel emulation~\cite{NetEquiv,NetEquiv2}, on the other hand, focuses on the set of jointly typical input-output pairs
and the difference between the empirical joint distribution (averaged over ensembles of channel emulators)
and the desired joint distribution is quantified by a multiplicative gap to ensure a small probability
of error events\footnote{Jointly typical pairs with  decoding error probability larger than a threshold~\cite{NetEquiv} are expurgated from channel emulators.}.
As we focus on characterizing capacity (bounds) rather
 than reconstructing (exact) common randomness for multi-terminal channels, we shall follow the channel emulation framework~\cite{NetEquiv, NetEquiv2} when constructing bounding models for BCs and MACs.

It is, however, non-trivial  to apply the network
equivalence tools~\cite{NetEquiv,NetEquiv2,Effros1,Effros2} onto
wireless networks owing to the broadcast nature of wireless
transmission. On one hand, the bounding models proposed
in~\cite{NetEquiv2} for MACs/BCs with $m$ transmitters/receivers
contain up to $(2^m{-}1)$ bit pipes
,  leading to
computational  inefficiency when $m$ is large (as in a wireless
hot-spot which may contain potentially many users). On the other
hand, the received signal at a terminal may contain several
broadcasted signals, which creates dependence/interference among
several transmitter-receiver pairs. Although such dependence has
been partially incorporated into ICs, the whole family of
non-layered multi-hop channels (e.g., relay channels) have been excluded from
consideration since the channel emulation techniques are developed
for single-hop channels.

In this paper, we present simple but efficient methods to construct upper and lower bounding models for wireless
networks with potentially many nodes, at a complexity that grows linearly with the number of nodes.
We propose a channel decoupling approach to decompose a memoryless wireless network into
 decoupled BCs and MACs.
In our bounding models, the constraints on the sum rate and on the individual rates are characterized
by different channel emulation techniques as inspired by~\cite{FlavioITW}.
Our upper bounding models, which consist of only point-to-point bit pipes, are constructed by first extending  the one-shot models for MACs/BCs to
 many-user scenarios and then integrating them with the channel  emulation techniques. Our lower bounding
 models, which contain both  point-to-point and point-to-points (hyper-arc) bit pipes, are obtained based on a two-step update of the decoupled BCs and MACs by taking  their dependence into account.
The main advantage of our proposed bounding models are their simplicity and the fact that
they can be easily extended to large networks. We demonstrate by
 examples that the resulting upper and lower bounds can approach  the capacity in some setups.

  Throughout this paper, we assume memoryless independent sources as in~\cite{NetEquiv}, where the optimality of channel-network  separation
is established for networks with only point-to-point channels. Extension to memoryless correlated sources  can be found in~\cite{Tian14, Jalali11} where channel-network separation is established  in the context of lossy~\cite{Tian14, Jalali11} and lossless~\cite{Jalali11} source coding problems over networks with only point-to-point DMCs. Extension to AWGN channels is also established in~\cite{Jalali11}.
We also assume that the distortion components (e.g., noise) are independent from
the transmitted signals. This assumption can be relaxed in
scenarios where the noise power is dependent on the power of input
signals, and in such scenarios we take the smallest (resp.
largest) noise power when constructing upper  (resp. lower)
bounding models. We further assume that the distortion components
at receiving nodes within a coupled\footnote{The definition of
coupled and non-coupled networks will be introduced in
Sec.~\ref{sec:Eqv:def}.} BC are mutually independent\footnote{Though we still
allow noise correlation at receiving nodes within a non-coupled
BC.}, and the scenario of coupled BC with
correlated noise will be investigated in our future work.
   Note that in this paper our focus is on constructing
noiseless bounding networks that can serve as a basis to compute
capacity bounds, rather than finding the capacity of a
noiseless network, which itself is a very difficult
problem~\cite{Lehman2, Grant08}. We refer to~\cite{NC_acyc,
IT_network, sLNC, AlgebraNC, Edge-cut06, GNS, NetSharing06} for
various computational tools available to characterize (bounds on)
the capacity of noiseless networks.

There are some other methods aiming at characterizing the capacity of wireless networks.
 A deterministic approach
proposed in~\cite{ADT} can approximate the capacity of Gaussian
networks within a constant gap in the high signal-to-noise ratio
(SNR) regime, where amplify-and-forward (AF) has been proved to approach the capacity
in multi-hop layered networks~\cite{ANC12}. A layering approach with a global information
 routing proposed in~\cite{PhyFlow} for wireless
networks with non-coupled BCs and MACs can provide lower bounds
 within a multiplicative gap from the capacity upper
bound. Capacity approximations for multiple unicast transmissions over coupled wireless networks developed in \cite{Kannan14} combines polymatroidal network analysis with carefully designed coding schemes for each independent single-hop interference channel, and the approximation accuracy is characterized for  bidirectional networks with symmetric fading coefficients.  Since we aim at approaches that can be used in all SNR regions and for all communication tasks, we do not follow the methods developed in~\cite{ADT,ANC12, PhyFlow, Kannan14}.

The rest of this work is organized as follows.We first introduce in Sec.~\ref{sec:Eqv:background} a few important definitions and
a brief introduction of the network equivalence theory and the
one-shot method. Then in
Sec.~\ref{sec:Eqv:noncoup} we present our improvement on the
bounding models for independent BCs and MACs. In
Sec.~\ref{sec:Eqv:coup} we describe the network decoupling
method  for coupled networks and demonstrate how the upper and lower bounding
 models are constructed by taking the coupled structure into account.
We illustrate our bounding models in
 Sec.~\ref{sec:coupled:illust} by constructing upper and lower bounding models for coupled
 networks and  conclude this work in Sec.~\ref{sec:Eqv:conc}.

\section{Network Equivalence Theory and the One-Shot Bounding
Method}\label{sec:Eqv:background}

In this section we present a few basic definitions that are frequently used in our paper and give a brief introduction of the stacked as well as the one-shot channel emulation techniques. We inherent the setups for bounding models and channel emulation arguments from~\cite{NetEquiv, NetEquiv2}, but our notation is slightly different  to minimize the number of sub-/sup-scripts. $X$ represents a random variable drawn from an alphabet $\mX$ with cardinality $|\mX|$, and $x$ is a realization with probability $p(x)\triangleq p_X(X{=}x)$. We use subscripts ($X_i, x_i$) to differentiate random variables and use superscripts ($x^N$) to indicate the number of realizations drawn independently from the same random variable. $X_{[1:n]}$ refers to $\{X_1,X_2,\ldots,X_n\}$ and $x_{[1:n]}$ is the corresponding set of realizations, one from each random variable. We use $p(\by|\bx)$ to represent the transition function of a memoryless network where $\bx$ is a vector containing one realization from every channel input alphabet and $\by$ is the vector for all channel outputs. The dimension of $\bx$ and $\by$ depends on the specific network and therefore will not be specified unless necessary.

\subsection{Basic Definitions}\label{sec:Eqv:def}

We represent a memoryless channel/network by a triplet consisting of the input alphabets, the output alphabets, and a conditional probability distribution (i.e., the transition function) that can fully characterize its behavior\footnote{As in~\cite{NetEquiv, NetEquiv2}, we only focus on channels/networks where the transition functions exist.}. A point-to-point memoryless channel with input alphabet $\mX_i$, output alphabet
$\mY_j$, and transition probability $p(y_j|x_i)$ is therefore denoted by
\[\mbN_{i\to j}\triangleq (\mX_i, p(y_j|x_i), \mY_j).\]
In wireless networks, a node may receive signals from multiple transmitters via MAC or parallel channels, and/or transmit to multiple receivers via BC or parallel channels. Therefore a node may associate with multiple input/output alphabets. For a given node $v$, letting $I_v$ be the number of its incoming parallel channels and $O_v$ be the number of its outgoing parallel channels, we denote the alphabets associated with its outgoing channels by the Cartesian product $\mX_v \triangleq \prod_{i=1}^{O_v} \mX_{v,i}$ and the alphabets associated with its incoming channels by $\mY_v \triangleq \prod_{j=1}^{I_v} \mY_{v,j}$. A memoryless network $\mbN_T$ with the set of nodes $T$ and the transition function $p(\by|\bx)$  can be fully represented as
\begin{align}
\mbN_T & \triangleq  (\prod_{v\in T} \mX_v,  p(\by|\bx), \prod_{v\in T} \mY_v) \label{eqn:model:network1}\\
       &  = (\prod_{v\in T}\prod_{i=1}^{O_v} \mX_{v,i},  p(\by|\bx), \prod_{v\in T} \prod_{j=1}^{I_v} \mY_{v,j}) \label{eqn:model:network2}\\
       &  = (\prod_{n} \mX_n, p(\by|\bx),\prod_{m} \mY_m). \label{eqn:model:network3}
\end{align}
Note that the structure and behavior of a memoryless network are explicitly characterized by the triplet  in \eqref{eqn:model:network2}, which
 specifies the association of input/output alphabet(s) to a specific node in the network.
The triplet in \eqref{eqn:model:network3}, on the other hand, focuses on each individual channel/sub-network and highlights its associated input/output alphabets,  where the underlying network structure is implicitly assumed.
Unless necessary, hereafter we will focus on the model \eqref{eqn:model:network3} without specifying $\mX_n$ as the solo alphabet of a node or as one of its outgoing alphabets.

\begin{definition}[Independent Channel]\label{dfn:indchannel}
A point-to-point channel $\mbN_{i\to j} = (\mX_i,
p(y_j|x_i), \mY_j)$  within a memoryless network $\mbN_T =  (\prod_{n} \mX_n,
p(\by|\bx),\prod_{m} \mY_m)$ is said to be independent if the network transition function
$p(\by|\bx)$ can be partitioned as
\begin{align}
p(\by|\bx)= p(\by_{/j}|\bx_{/i})p(y_j|x_i),
\end{align}
where $\bx_{/i}$ denotes the vector of $\bx$ without element
$x_i$, and similarly for $\by_{/j}$.
\end{definition}

To highlight the independence, we emphasize the notation for network
$\mbN_T$ as
\begin{align}
\mbN_T& =\mbN_{i\to j} \times \mbN_{T/ (i{\to}j)}\\
&  \triangleq (\mX_i,
p(y_j|x_i), \mY_j) \times (\prod_{n\neq i} \mX_n, p(\by_{/j}|\bx_{/i}), \prod_{m\neq j} \mY_m).\nonumber
\end{align}
A sub-network $\mbN_{S\to D} {\triangleq}  (\prod_{n\in S} \mX_n, p(\by_D|\bx_S), \prod_{m\in D} \mY_m)$
within $\mbN_T$ is said to be independent, denoted by
\begin{equation}
\mbN_T = \mbN_{S\to D} \times \mbN_{T/ (S{\to}D)},
\end{equation}
 if the network transition probability can be partitioned as
\begin{align}\label{eqn:indp:network}
p(\by|\bx)= p(\by_{D}|\bx_{S}) p(\by_{/D}|\bx_{/S}).
\end{align}
An independent sub-network may be further partitioned into several independent channels and/or sub-networks.

\begin{definition}[Coupled/Non-coupled Network]\label{dfn:Ncoupled}
A network is said to be coupled if any of its channels is part of a MAC and a BC simultaneously. That is, it contains a sub-network $\mbN_{S\to D}$ with $|S|\geq 2$, $|D|\geq 2$, and its transition function $p(\by_{D}|\bx_{S})$  can't be partitioned into non-trivial product format.
Otherwise, the network is non-coupled. A MAC and a BC are said to be coupled if they share a common link.
\end{definition}

For example, the classical three-node relay channel $p(y,y_1|x,x_1)$ is coupled and the two-hop diamond network is non-coupled. Wireless networks,  as expected, are in general coupled owing to the broadcast nature of microwave propagation.

\begin{definition}[Bit pipe]\label{dfn:bit-pipe}
A point-to-point bit pipe of rate  $R\geq 0$ is a noiseless channel that can reliably transmit $\lfloor nR\rfloor$ bits during $n$ channel uses for any positive integer $n$. It is represented as
\[\mbC_{i\to j} = (\{0,1\}^{\lfloor nR\rfloor}, \delta(y_j - x_i), \{0,1\}^{\lfloor nR\rfloor}), \forall n\geq 1,\]
where $\delta(\cdot)$ is the Kronecker delta function. A point-to-points bit pipe (hyper-arc) is denoted by
  \[\mbC_{i\to J} = (\{0,1\}^{\lfloor nR\rfloor}, \prod_{j\in J}\delta(y_j - x_i), \prod_{j\in J}\{0,1\}^{\lfloor nR\rfloor}), \forall n\geq 1,\]
  where $|J|$ is the number of heads of this hyper-arc.
\end{definition}

\begin{definition}[Capacity Bounding Models~\cite{NetEquiv}]\label{dfn:Cbound}
Given two independent (multi-terminal) channels  $\mbC$ and $\mbN$,
$\mbC$ is said to upper bound $\mbN$, or
equivalently $\mbN$ lower bounds $\mbC$, if the
capacity (region) of $\mbN\times \mbW$ is a subset
of that for $\mbC\times \mbW$ for any network
$\mbW$. We denote their relationship by $\mbN
\subseteq \mbC$. $\mbC$ and $\mbN$ are said
to be equivalent if $\mbC \subseteq \mbN \subseteq
\mbC$.
\end{definition}

For an independent noisy channel $\mbN$, we
construct channels  $\mbC_u$ and $\mbC_l$
consisting of only noiseless bit pipes, such that $\mbC_u$
is the upper bounding model and $\mbC_l$ is the lower
bounding model for $\mbN$, i.e.,
\begin{align}
\mbC_l \subseteq \mbN \subseteq \mbC_u.
\end{align}

\subsection{Network Equivalence Theory for Independent Channels}

\begin{figure}[t]
\centering
\psfrag{xn}[][]{$\mX^N$}\psfrag{yn}[][]{$\mY^N$}
\psfrag{x}{$\mX$}\psfrag{y}[][]{$\mY$}
\psfrag{pxy}[][]{$\ \ p(y|x)$} \psfrag{ppxy}[][]{$\prod_{i=1}^{N} p(y_i|x_i)$}
\psfrag{eq}[][]{$ \Leftrightarrow$}\psfrag{eq2}[][]{$\longleftrightarrow$}
\psfrag{XN}[][]{$x^{N}$}\psfrag{YN}[][]{$y^{N}$}
\psfrag{0n}[][]{$W$}\psfrag{1n}[][]{$\hat{W}$}
\psfrag{af}[][]{$\alpha(\cdot)$}\psfrag{bf}[][]{$\alpha^{-1}(\cdot)$}
\psfrag{en}[][]{$\beta(\cdot)$}\psfrag{de}[][]{$\beta^{-1}(\cdot)$}
\psfrag{qq}[][]{$\{0,1\}^{NR}$}\psfrag{q1}[][]{$\{0,1\}^{R}$}
\psfrag{C}{$\mbN$}\psfrag{R}{$\mbC$}
  \includegraphics[width=8.5cm]{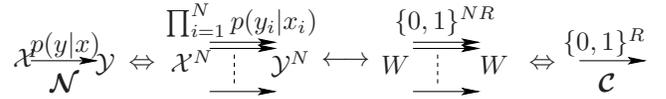}\\
  (a) Capacity regions of a network and its stacked network are identical\\
  \includegraphics[width=8.5cm]{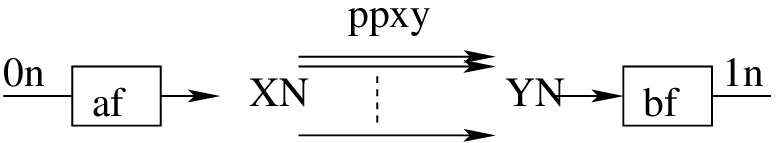}\\
  (b) Channel coding argument to prove $\mbC {\subseteq} \mbN$ for all  $R{<}C$
    \includegraphics[width=8.5cm]{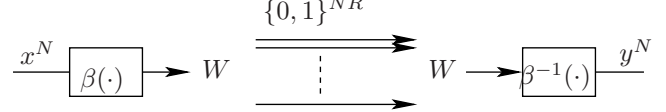}\\
  (c) Lossy source coding argument to prove $\mbN {\subseteq} \mbC$ for all $R{>}C$
    \caption{A point-to-point noisy channel $\mbN=(\mX, p(y|x),\mY)$ with capacity $C=\max_{p(x)} I(X;Y)$
     and a noiseless point-to-point bit pipe $\mbC$ of rate $R$ are said to be equivalent if $R=C$, where the equality
     comes from the continuity of the capacity region. The input/output of their corresponding
     stacked networks are $x^{N}\in\mX^N$, $y^{N}\in\mY^N$, and $W, \hat{W} \in\{1,\ldots, 2^{NR}\}$.}
    \label{fig:Neq:model}
\end{figure}

In~\cite{NetEquiv}, the equivalence between an independent
point-to-point noisy channel $\mbN$  and
a noiseless point-to-point bit pipe $\mbC$ has been established, as long as the with capacity of  $\mbN$ equals the rate of $\mbC$, by showing that any code that runs reliably
on a network $\mbN\times \mbW$ can also operate on
$\mbC\times \mbW$ with asymptotically vanishing
error probability. The  argument is based on a channel emulation technique over a stacked network
where $N$ parallel replicas
of the network have been put together to run the code. As
illustrated in Fig.~\ref{fig:Neq:model}, for  noisy $\mbN$ with capacity $C$ and
rate-$R$ bit pipe $\mbC$, the proof can be divided
into three steps.

In Step I, a network and its $N$-fold stacked network (consisting
of $N$ parallel replicas of the network) are proved to share the
same rate region by showing that any code that can run on the
network can also run on its stacked network, and vice versa.
Therefore we only need to show the equivalence between the stacked
network for $\mbN$ and that for $\mbC$, as
illustrated in Fig.~\ref{fig:Neq:model}(a).

In Step II, the proof of $\mbC \subseteq \mbN$
employs a channel coding argument over the stack of $N$ channel
replicas as illustrated in Fig.~\ref{fig:Neq:model}(b):  A message
$W$ of $2^{NR}$ bits is mapped by the channel encoder
$\alpha(\cdot)$ onto a codeword $x^{N}$ of length $N$, and then
transmitted over the $N$-stack noisy channels, with one symbol on
each replica, such that reliable transmission over the noisy
stacked network can be realized with arbitrary small error
probability as $N$ goes to infinity for all $R<C$.

In Step III, the proof of $\mbN \subseteq \mbC$ is
based on a lossy source coding argument as illustrated in
Fig.~\ref{fig:Neq:model}(c): The input sequence $x^{N}$ to the
noisy stacked network is first quantized/compressed by a lossy
source encoder $\beta(\cdot)$ into $2^{NR}$ bits, represented by
the message $W$, which is then transmitted through the noiseless
stacked network. The reconstructed sequence $y^{N}$ is
selected in such a way that it is jointly typical  with the
transmitted sequence  $x^{N}$, in contrast to the usual
distortion measure. The existence of a good lossy source coding
codebook for any $R>C$ is proved by a random coding argument,
i.e., by showing that the average error probability over the
randomly chosen ensemble of codebooks is small.

Finally, the equivalence between $\mbN$ of capacity $C$ and
$\mbC$ of rate $R$ can be established when $R=C$
based on the continuity of the capacity region. Readers are referred to~\cite{NetEquiv} for a rigorous and thorough treatment.

Note that the jointly typical pairs $(x^{N}, y^{N})$ that are used to construct the channel emulation codebooks are taken from a ``restricted'' typical set $\hat{A}^{(N)}_{\epsilon}$ where the associated decoding error probability (assuming $x^{N}$ is transmitted through the original noisy channel) is smaller than a threshold. Sequences that do not satisfy this condition are expurgated. That is, given the classical typical set
\[T^{(N)}_{\epsilon} = \{(x^N,y^N)| |\frac{1}{N}\log(p(x^N,y^N))-H(X,Y)|<\epsilon\},\]
the restricted typical set is defined~\cite{NetEquiv} as
\[\hat{A}^{(N)}_{\epsilon} = \{(x^N,y^N)| (x^N,y^N)\in T^{(N)}_{\epsilon}, p(T^{(N)}_{\epsilon}|x^N)>\frac{1}{2}\}.\]

The concept of capacity upper and lower bounding models developed
in~\cite{NetEquiv} has been extended to independent multi-terminal
channels in~\cite{NetEquiv2} following  similar arguments as
illustrated in Fig.~\ref{fig:Neq:model}, and multiplicative  and
additive gaps between lower and upper bounding models for
independent multi-terminal channels have been established.
Illustrative upper and lower bounding models for MACs/BCs/ICs
involving two transmitters and/or two receivers have been
demonstrated in~\cite{NetEquiv2,Effros1,Effros2}. Given a noisy
network composed by independent building blocks whose upper and
lower bounding models are available, we can replace these building
blocks with their corresponding upper (resp. lower) bounding models and
then characterize an outer (resp. inner) bound for its capacity region
based on the resulting noiseless network models.

For wireless networks, however, it may be difficult in general to apply directly the channel emulation technique to
construct  bounding models, as the coupled components may involve many transmitting/receiving
nodes. For coupled single-hop networks which can be modeled as
ICs, the bounding models are difficult to characterize even for the
simplest $2 \times 2$ setup~\cite{NetEquiv2}.  For coupled
multi-hop non-layered networks, it is unclear how the channel emulation
technique can be extended to incorporate the interaction among different transmitting-receiving pairs across different layers. Although one may apply the cut-set bound~\cite{CoverThomas}
to construct upper bounds for wireless networks, the resulting analysis
may become quite involved, as illustrated in~\cite{Cover79, NC_relay,
NC_X_relay, JSAC11}, for characterizing upper bounds for small
size relay networks. Moreover, even if we manage to construct
bounding models for a specific coupled network, we have to create
new bounding models for each different network topology, which
makes it unscalable for wireless networks that have diversified
communication scenarios and topologies.

\subsection{One-shot Bounding Models}

\begin{figure}[t]
\centering
\psfrag{x1}[][]{$\mX_1$}\psfrag{x2}[][]{$\mX_2$}\psfrag{xn}[][]{$\mX_m$}
\psfrag{y1}[][]{$\mY_1$}\psfrag{y2}[][]{$\mY_2$}\psfrag{yn}[][]{$\mY_m$}
\psfrag{x}{$\mX$}\psfrag{y}[][]{$\mY$}
\psfrag{l1}[][]{$l_1$}\psfrag{l2}[][]{$l_2$}\psfrag{ln}[][]{$l_m$}\psfrag{ls}[][]{$l_s$}
 \psfrag{ni}{$n_I$}\psfrag{di}{$n_I$}
   \includegraphics[width=8.5cm]{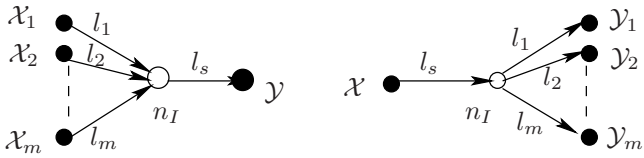}
    \caption{The one-shot upper bounding models for MACs/BCs with $m$ transmitters/receivers. The white nodes indicated by $n_I$ are auxiliary operation nodes    to specify the rate constraints on the sum rate and on individual rates. All the channels are noiseless bit pipes
    and independent from others. The one-shot upper bounding
    models are fully characterized by the rate vector $(R_{l_s}, R_{l_1}, \ldots, R_{l_m})$,
    where $R_{l_i}$ is the rate of the noiseless bit pipe $l_i$.}
    \label{fig:Neq:oneshot}
\end{figure}

\begin{figure}[t]
\centering
\psfrag{x1}[][]{$\mX_1$}\psfrag{x2}[][]{$\mX_2$}\psfrag{xn}[][]{$\mX_m$}\psfrag{x}[][]{$\mX$}
\psfrag{y1}[][]{$\mY_1$}\psfrag{y2}[][]{$\mY_2$}\psfrag{y}[][]{$\mY$}
\psfrag{l1}[][]{$\mX_1$}\psfrag{ln}[][]{$\mX_m$}\psfrag{ls}[][]{$\mX'$}
\psfrag{d1}[][]{$\mY_1$}\psfrag{d2}[][]{$\mY_2$}\psfrag{ds}[][]{$\mY'$}
\psfrag{z1}[][]{$\mZ_1$}\psfrag{zn}[][]{$\mZ_m$}\psfrag{zs}[][]{$\mY$}
\psfrag{v1}[][]{$\mX$}\psfrag{v2}[][]{$\mX$}\psfrag{vs}[][]{$\mX$}
  \includegraphics[width=8.5cm]{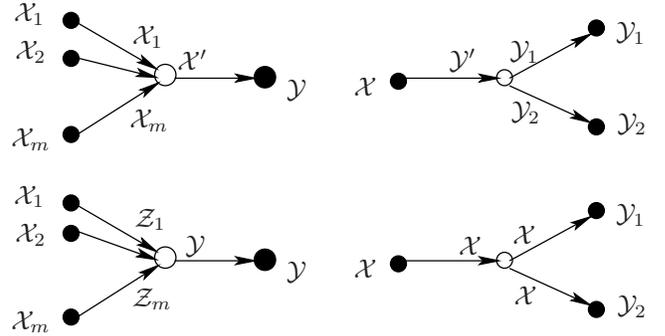}
    \caption{Illustration of the two types one-shot upper bounding models~\cite{FlavioITW} for $m$-user MACs  and for two-user BCs: $\mbC_{u, MAC, 1}$ (upper left), $\mbC_{u, MAC, 2}$ (lower left), $\mbC_{u, BC, 1}$ (upper right),  and  $\mbC_{u, BC, 2}$ (lower right). The operation of the auxiliary operation nodes  are explicitly specified by the associating input/output alphabets, where $\mX'=\prod_{i=1}^m \mX_i$ and $\mY'=\mY_1\times\mY_2$ are compound  input/output alphabets, and $Z_i\in\mZ_i, i=1,\ldots,m$ are auxiliary random variables to synthesize $Y\in\mY$ through a predefined function $y=g(z_1,\ldots,z_m)$ as described in \eqref{eqn_Py_mac2}.}
    \label{fig:Neq:oneshot-models}
\end{figure}

Instead of using emulation with channel coding or lossy source
coding to construct bounding models as in~\cite{NetEquiv,
NetEquiv2}, a class of one-shot upper bounding models have been proposed
in~\cite{FlavioITW} for independent MACs/BCs, where an
 auxiliary operation node is introduced for each MAC/BC to facilitate separate characterization of  the sum rate and the individual rates.
  As illustrated in Fig.~\ref{fig:Neq:oneshot}, all the channels in the one-shot upper bounding models are  point-to-point bit pipes
  which can be fully characterized by the rate vector $(R_{l_s}, R_{l_1}, \ldots, R_{l_m})$, where $R_{l_i}$ is the rate of the bit pipe $l_i$.
     The rate constraint $R_{l_i}$ can either be constructed by channel emulation over  the stacked network as in~\cite{NetEquiv, NetEquiv2}, or by channel emulation that is realized in each instance corresponding to a
channel use (hence referred to as ``one-shot'' approach).  For example, we can construct the sum rate constraint by stacked emulation
whilst bound individual rates by one-shot emulation, or the other way around, which results in two types of upper bounding models~\cite{FlavioITW}. To highlight the specific channel emulation method used on each link, we illustrate in Fig.~\ref{fig:Neq:oneshot-models} the explicit operation of the auxiliary
node by specifying its input/output alphabets.

\subsubsection{One-Shot Channel Emulation for Individual Rate Constraints}

For MACs with $m$ transmitters, say  $\mbN_{MAC} = (\prod_{i=1}^m \mX_i, p(y|\bx), \mY)$,
the one-shot emulation for the channel from transmitter with alphabet
$\mX_i$ sends exactly the corresponding source symbol $X_i\in\mX_i$ to the auxiliary operation
node $n_I$, hence requiring a noiseless bit pipe of rate $R_{l_i}\geq
\log(|\mX_i|)$.  The auxiliary node then combines all the inputs together and formulates a super alphabet
 $\mX'=\prod_{i=1}^m \mX_i$. The channel from $n_I$ to the receiver is then simply a point-to-point channel with input alphabet $\mX'$ and output alphabet $\mY$. This setup is illustrated in Fig.~\ref{fig:Neq:oneshot-models} (upper left).
 By the network equivalence
theory for point-to-point channels~\cite{NetEquiv}, successful channel
emulation requires
\begin{align}
R_{l_s} \geq R_{MAC} {\triangleq} \max_{p(x')} I(X'; Y) {=} \max_{p(x_1,\ldots,x_m)} I(X_1,\ldots, X_m; Y).
\end{align}
Hence, we can construct the upper bounding model  as
\begin{align}\label{eqn:one:MAC1}
\mbC_{u, MAC, 1} = (R_{MAC}, \log(|\mX_1|), \ldots, \log(|\mX_m|)).
\end{align}

Similarly, for a BC with two receivers $\mbN_{BC} = (\mX, p(y_1,y_2|x), \mY_1\times\mY_2)$,
the channel from the source to the auxiliary operation node $n_I$ has a super output alphabet $\mY'=\mY_1\times\mY_2$,
which results in a sum rate constraint
\begin{align}
R_{BC} \triangleq \max_{p(x)} I(X; Y') =  \max_{p(x)} I(X; Y_1,Y_2).
\end{align}
After successfully receiving $Y' = [Y_1, Y_2]$, the  auxiliary   node $n_I$ sends exactly the corresponding symbol $Y_i$ to the receiver  with alphabet $\mY_i$,  hence requiring a noiseless bit pipe of rate $R_{l_i}\geq
\log(|\mY_i|)$. This setup is illustrated in Fig.~\ref{fig:Neq:oneshot-models} (upper right). The corresponding upper bounding model is therefore
\begin{align}
\mbC_{u, BC, 1} = (R_{BC}, \log(|\mY_1|), \log(|\mY_2|)).
\end{align}

\subsubsection{One-Shot Channel Emulation for Sum Rate Constraint}

Alternatively, one may bound the sum rate by one-shot channel emulation and then specify individual rates.
For the two-user BC $\mbN_{BC}$, the one-shot emulation sends every $X$ from the source to the auxiliary node, which requires a sum rate constraint of $\log(|\mX|)$. Then the auxiliary node sends $X$ to the two receivers through independent channels, each requires a rate constraint
\begin{align}
R_{i} \triangleq \max_{p(x)} I(X; Y_i), i=1, 2.
\end{align}
This setup is illustrated in Fig.~\ref{fig:Neq:oneshot-models} (lower right) and the corresponding upper bounding model is
\begin{align}
\mbC_{u, BC, 2} = (\log(|\mX|), R_1, R_2),
\end{align}
which is valid only if the noise at two receivers are independent, i.e., the transition probability
can be factorized as
\[p(y_1,y_2|x)=p(y_1|x)p(y_2|x).\]

For the $m$-user MAC $\mbN_{MAC}$, the setup for channel emulation is illustrated in Fig.~\ref{fig:Neq:oneshot-models} (lower left) where $z_i\in\mZ_i$ is an auxiliary random variable for the channel from $\mX_i$  such that~\cite{FlavioITW}
\begin{align}\label{eqn_Py_mac2}
p(y|x_1,\ldots,x_m)= \sum_{\substack{z_1,\ldots,z_m:\\ y=g(z_1,\ldots,z_m)}}\prod_{i=1}^m p(z_i|x_i),
\end{align}
where $g:\prod_i \mZ_i \to \mY$ is a predefined deterministic function
\begin{align}
y = g(z_1, z_2, \ldots, z_m),
\end{align}
to reproduce the channel output $Y$ at the auxiliary node $n_I$.
From stacked channel emulation we can get the individual rate constraints as follows
\begin{align}
R_{i} \triangleq \max_{p(x_i)} I(X_i; Z_i), i=1,\ldots,m.
\end{align}
Therefore the corresponding one-shot model can be written as
\begin{align}\label{eqn:one:MAC2}
\mbC_{u, MAC, 2} = (\log(|\mY|), R_1, \ldots, R_m).
\end{align}

Here we give two examples to show how to construct the auxiliary random variables as specified by \eqref{eqn_Py_mac2}.
For Gaussian MACs, auxiliary random variables can be constructed  based on a noise partitioning
approach, i.e., the additive noise at the destination is
partitioned into  independent parts and allocated to each of the individual channels.
 For a two-user Gaussian MAC with  noise power $\sigma^2{=}1$ and the received power constraint $\gamma_i$ for $X_i$,
  the corresponding upper bounding model is
\begin{align}\label{eqn:partition:add}
\mbC_{u, MAC, 2} = \left(\log(|\mY|), \frac{1}{2}\log\left(1{+}\frac{\gamma_1}{\alpha}\right),
\frac{1}{2}\log\left(1 {+}\frac{\gamma_2}{1-\alpha}\right)\right),
\end{align}
where $\alpha\in(0,1)$ is the noise
 partitioning parameter chosen to minimize the total input
rate
\begin{align}
R_{s} = \frac{1}{2}\log\left(1+\frac{\gamma_1}{\alpha}\right)+\frac{1}{2}\log\left(1+\frac{\gamma_2}{1-\alpha}\right).
\end{align}
%
 For binary additive MAC $\{0,1\}^2\to\{0,1\}$ with Bernoulli distortion $Bern(\epsilon)$, the corresponding
distortion parameter $\epsilon_i$ for channel $l_i$ should satisfy
\begin{align}\label{eqn:partition:bin}
\epsilon = \epsilon_1(1-\epsilon_2) + \epsilon_2(1-\epsilon_1).
\end{align}

\begin{remark}
Although $\mbC_{u, MAC, 1}$ and $\mbC_{u, BC, 1}$ are tight on sum rate in the
sense that there are some kind of networks where the sum rate
constraint $R_{MAC}$ ($R_{BC}$) is tight, the constraints on individual rates
are somewhat loose. $\mbC_{u, MAC, 2}$ and $\mbC_{u, BC, 2}$, on the other hand, have tighter bounds on individual rates but looser on sum rate.
\end{remark}

\subsubsection{Gap in One-shot Bounding Models}

The gap between the upper and lower bounding models for Gaussian
channels and for binary symmetric channels have been examined
in~\cite{FlavioITW}, where a gap less than $1/2$ bit per channel
use has been established for MACs with two transmitters and BCs
with two receivers.

\section{Bounding Models for Non-coupled Networks}\label{sec:Eqv:noncoup}

 For non-coupled networks, which can be decomposed into
independent MACs/BCs and point-to-point channels,  we first
construct upper and lower bounding models for MACs/BCs, which can then be used to substitute
their noisy counterparts in construction of noiseless bounding models for the original noisy network.
To give a full
description of all the rate constraints on any subset of transmitters/receivers,
the upper and lower bounding models for independent MACs/BCs with  $m$ transmitters/receivers need to
consist of $(2^m{-}1)$ rate constraints. However, such an approach is  not scalable as $m$ can be quite
large in many practical scenarios. Instead, we introduce a rate vector of
length up to $(m{+}1)$ to specify our upper and lower bounding models.


For independent MACs/BCs with  $m$ transmitters/receivers, our
upper bounding models inherent the one-shot model structure as illustrated in
Fig.~\ref{fig:Neq:oneshot} and therefore only contain constraints on  each of the
maximum allowed individual rate $R_i$, $i=1,\ldots,m$, and the
total sum rate. All the constraints on subsets
of individual rates, i.e.,
\begin{align}
R(S)\triangleq\sum_{i\in S}R_i, \ S\subset \{1,\ldots,m\} \mbox{, and } 2 \leq |S| < m,
\end{align}
 are omitted, which results in a looser but simpler
upper bound.
  The benefits of keeping the one-shot structure are two
fold: they can be extended to
MACs/BCs  with $m$ transmitters/receivers at low complexity; they
facilitate our proposed channel decoupling method in a natural way
for constructing the upper and lower bounding bounds for coupled networks.

\subsection{Upper Bounding Models for Independent MACs with $m$ Transmitters}\label{sec:noncoup:MAC:upper}

For MACs with $m$ transmitters, the one-shot bounding model $\mbC_{u, MAC, 1}$ defined in \eqref{eqn:one:MAC1} focuses solely  on the sum rate and $\mbC_{u, MAC, 2}$ defined in \eqref{eqn:one:MAC2} focuses solely on individual rates.
To facilitate a tradeoff between the sum rate constraint and  each of the individual rate constraints,
we propose here a new upper bounding model,
\begin{align}
\mbC_{u, MAC, new}& = (R_s, R_1, \ldots, R_m),\label{eqn:MACnew}\\
 R_s & = \max_{p(v_1,\ldots,v_m)} I(V_1\ldots V_m;Y),\nonumber\\ 
 R_i & = \max_{p(x_i)} I(X_i;V_i), \ i=1,\ldots,m, \nonumber 
\end{align}
where $V_1,\ldots,V_m$ are auxiliary random variables introduced to account for the noise ``allocation'' to each individual rate constraint through $V_i$ and its conditional probability $p(v_i|x_i) $ such that
\begin{align}
p(v_1,\ldots,v_m|x_1,\ldots,x_m)  =  \prod_{i=1}^m p(v_i|x_i). \label{eqn:pVX}
\end{align}
Noise allocation to the sum rate constrain is through $V_1,\ldots,V_m$ and the conditional probability $p(y|v_{[1:m]})$ such that
\begin{align}
p(y,v_1,\ldots,v_m|x_1,\ldots,x_m)& =  p(y|v_{[1:m]}) p(v_{[1:m]}|x_{[1:m]}), \label{eqn:pYVX}
\end{align}
i.e., $Y$---$V_{[1:m]}$---$X_{[1:m]}$ forms a Markov chain.

For the new upper bounding model $\mbC_{u, MAC, new}$, we introduce a parameter $\alpha\in[0,1]$ to quantify the proportion of noise distortion that has been ``allocated'' to the sum rate constraint $R_s$. It is determined by the type of the channel/noise via its channel transition function $p(y|x_{[1:m]})$ and by the auxiliary random variables $V_1,\ldots, V_m$ via the conditional probability function $p(y|v_{[1:m]})$. When all the noise distortion is put into $R_s$, i.e., $V_i=X_i, \forall i$ and thus $p(y|v_{[1:m]})=p(y|x_{[1:m]})$, we have $\alpha{=}1$ and the corresponding sum rate constraint $R_s$ is the tightest. When no noise is put into $R_s$, i.e., when $Y$ is fully determined by $V_1,\ldots, V_m$ and thus $H(Y|V_1\ldots V_m)=0$, we have $\alpha{=}0$ and the sum rate constraint is relaxed to the one-shot upper bound $\log(|\mY|)$. The explicit dependence of $\alpha\in[0,1]$ and the noise allocation is determined by the type of the noise.
For example, for additive noise with average power constraint, we may use the proportion of noise power to evaluate $\alpha$. Another option is to define $\alpha$ as the ration between $H(Y|V_1\ldots V_m)$ and $H(Y|X_1\ldots X_m)$, provided that the latter is finite.

\begin{remark}
The parameterized bounding model $\mbC_{u, MAC, new}$  includes the two one-shot models as special cases:
putting all distortion into $R_s$ ($\alpha{=}1$) will generate $\mbC_{u, MAC, 1}$, which give us a tighter bound $R_{MAC}$ on the sum rate but looser constraints on all individual rates;
taking no noise into $R_s$ ($\alpha{=}0$) will produce $\mbC_{u, MAC, 2}$, which leads to looser bound on the sum rate but tighter bounds on individual rates.
\end{remark}

\subsubsection{$2$-user MACs}

\begin{figure}[t]
\centering
\psfrag{x1}[][]{$\mX_1$}\psfrag{x2}[][]{$\mX_2$}\psfrag{y}[][]{$\mY$}
\psfrag{l1}[][]{$I(X_1;V_1)$}\psfrag{l2}[][]{$I(X_2;V_2)$}\psfrag{ls}[][]{$I(V_1,V_2;Y)$}
\psfrag{d1}[][]{$\log(|\mX_1|)$}\psfrag{d2}[][]{$\log(|\mX_2|)$}\psfrag{ds}[][]{$I(X_1,X_2;Y|U)$}\psfrag{di}[][]{$I(X_1;U)$}
 \includegraphics[width=8.6cm]{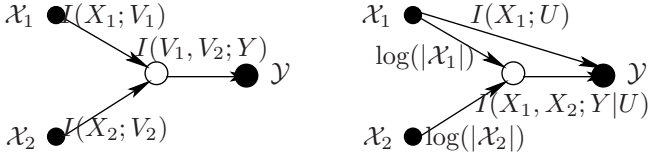}
    \caption{Upper bounding models for for independent MACs with two transmitters: the new bounding model $\mbC_{u, MAC, new}(\alpha)$ (left) and the model developed in~\cite[Theorem~6]{NetEquiv2}  with $p(u,y|x_1,x_2)=p(u|x_1)p(y|x_1,x_2)$ (right).  The label on each bit pipe is the minimum rate requirement for any given $p(x_1,x_2)$.}
    \label{fig:Neq:MAC2vs}
\end{figure}

For $2$-user MACs, it is interesting to compare the new upper bounding model $\mbC_{u, MAC, new}$ with the
 model developed in~\cite[Theorem~6]{NetEquiv2}, as illustrated in Fig.~\ref{fig:Neq:MAC2vs}.
On one hand, setting $V_1{=}X_1$ and $V_2{=}X_2$ will put all the distortion components into the sum rate constraint (i.e., $\alpha=1$), which results in an upper bounding model
\begin{align}\label{eqn:MAC2:a1}
\mbC_{u, MAC, new}&(\alpha{=}1) \\
&= (\max_{p(x_1x_2)} I(X_1,X_2; Y), \log(|\mX_1|),  \log(|\mX_2|)). \nonumber
\end{align}
It is the same as the upper bounding model in~\cite[Theorem~6]{NetEquiv2} when $U{=}\emptyset$.
On the other hand, if we choose $V_1{=}U$ and a deterministic function $f:\mU{\times}\mV_2\to\mY$ such that $y{=}f(u,v_2)$ and $p(u,v_2|x_1,x_2){=}p(u|x_1)p(v_2|x_2)$, we can generate another bounding model
\begin{align}\label{eqn:MAC2:a0}
\mbC_{u, MAC, new}&(\alpha{=}0) \\
& = (\log(|\mY|), \max_{p(x_1)}I(X_1;U), \max_{p(x_2)}I(X_2;V_2)).\nonumber
\end{align}
Compared to the bound in~\cite[Theorem~6]{NetEquiv2}, it is clear that $\mbC_{u, MAC, new}(\alpha{=}0)$ has a tighter bound on rate $R_1$.
However, as we will show  below, $\mbC_{u, MAC, new}(\alpha{=}0)$ has a looser bound on rate $R_2$.

\begin{lemma}\label{lemma:worseR2}
 Given $p(u,v_2|x_1,x_2){=}p(u|x_1)p(v_2|x_2)$ and $y{=}f(u,v_2)$,  we have
 \begin{align}
 I(X_1,X_2;Y|U)  \leq \min\{I(X_2;V_2), \log(|\mX_2|), \log(|\mY|)\}. \label{eqn:worseR2}
 \end{align}
\end{lemma}
\begin{IEEEproof}
\begin{align}
I(X_1,X_2;Y|U) & = H(Y|U) - H(Y|X_1,X_2,U) \label{eqn:Eqv:MAC2vs:r}\\
              & = H(V_2|U) - H(V_2|X_2)  \nonumber\\
              & \leq H(V_2) - H(V_2|X_2)\nonumber\\
              & = I(X_2;V_2) \leq \log(|\mX_2|),\nonumber
\end{align}
where the second equality is due to  $y{=}f(u,v_2)$ and the fact that $U$---$(X_1, X_2)$---$V_2$ and $X_1$---$X_2$---$V_2$ are Markov chains,
and the first inequality  comes from the fact that condition reduces entropy, with equality if and only if $X_1, X_2$ (and thus $U, V_2$) are independent. On the other hand, we can also rewrite  \eqref{eqn:Eqv:MAC2vs:r} as
\begin{align}
0 \leq I(X_1,X_2;Y|U) &\leq H(Y) - H(Y|X_1,X_2,U) \nonumber\\
  &= I(Y;X_1,X_2,U) \leq \log(|\mY|). \label{eqn:Eqv:MAC2vs:rs}
\end{align}
Combining \eqref{eqn:Eqv:MAC2vs:r} and \eqref{eqn:Eqv:MAC2vs:rs},  we have \eqref{eqn:worseR2}.
\end{IEEEproof}

Furthermore, if we are only interested in a tight bound on sum
rate, by Lemma~\ref{lemma:betterRs} below, we can see that choosing
a non-trivial auxiliary random variable $U$ in~\cite[Theorem~6]{NetEquiv2} can not improve the
sum rate constraint.

\begin{lemma}\label{lemma:betterRs}
Given $p(u,x_1,x_2,y)=p(u|x_1)p(x_1,x_2)p(y|x_1,x_2)$, we have
\begin{align}
I(X_1;U) + I(X_1,X_2;Y|U) \geq I(X_1,X_2;Y),
\end{align}
with equality if and only if $I(X_1,X_2;U|Y)=0$.
\end{lemma}
\begin{IEEEproof}
\begin{align*}
            & I(X_1,X_2;Y|U)\\
            & = I(X_1,X_2;Y) + I(X_1,X_2;U|Y) - I(X_1,X_2;U)\\
            & = I(X_1,X_2;Y) {+} I(X_1,X_2;U|Y) {-} I(X_1;U) {-}I(X_2;U|X_1)\\
            & = I(X_1,X_2;Y) + I(X_1,X_2;U|Y) - I(X_1;U)\\
            & \geq I(X_1 ,X_2;Y) - I(X_1;U), 
\end{align*}
where the first two equalities are due to the chain rule, the third equality comes from the
Markov chain $U$---$X_1$---$X_2$ (thus $I(X_2;U|X_1)=0$), and the last in equality comes from the fact that mutual information is non-negative,
with equality if and only if $I(X_1,X_2;U|Y)=0$.
\end{IEEEproof}

\subsubsection{$m$-user Gaussian MAC}

Given noise power $\sigma^2{=}1$ and the received power constraint $\gamma_i$ for $X_i$, $i{=}1,\ldots,m$,
the sum rate is upper bounded by
\begin{equation}\label{eqn:MAC:upp:sum}
R_{MAC}= \frac{1}{2}\log\left(1 + \left(\sum_{i=1}^m \sqrt{\gamma_i}\right)^2\right),
\end{equation}
which can be achieved only if all the transmitters can fully cooperate.

The parameterized new upper bounding model $\mbC_{u, MAC, new}(\alpha)$ defined in  \eqref{eqn:MACnew} can be constructed as follows.
 Let $Z_i$, $i=0,\ldots,m$, be independent Gaussian random variable  with zeros mean and variance $\alpha_i{>}0$ such that $\alpha_0=\alpha$ and $\sum_{i=0}^m  \alpha_i{=}1$. By  choosing the auxiliary random variables $V_i = X_i + Z_i$, $i=1,\ldots,m$, we have
\begin{align}
Y = V_1 + \ldots + V_m + Z_0,
\end{align}
and all the rate constraints in $\mbC_{u, MAC, new}(\alpha)$ can be written as
\begin{align}
& R_s(\alpha)  =\frac{1}{2}\log\left(1 + \frac{\left(\sum_{i=1}^m \sqrt{\gamma_i}\right)^2 + 1-\alpha}{\alpha}\right),\label{eqn:MAC:new:sum}\\
& R_i(\alpha)  = \frac{1}{2}\log\left(1 + \frac{\gamma_i}{\alpha_i} \right), \ i=1,\ldots,m.  \label{eqn:MACRs:alpha}
\end{align}

\begin{remark}
If we are interested in a tighter bound on any individual rate, say on
$R_{li}$, we can partition the noise by setting $\alpha=0$ and
$\alpha_i=1$, which leads to a tighter constraint
$R_i=\frac{1}{2}\log(1+\gamma_i)$ on rate $R_{li}$ but unbounded
constraints on all other individual rates and the sum rate. If we
are only interested in a tighter bound on the sum rate, setting
$\alpha=1$ will give us a tight bound $R_{MAC}$ on the sum rate
but unbounded constraints on all individual rates.
\end{remark}

 For given $\alpha\in[0,1)$, one way to determine the noise
partitioning parameters $\alpha_i$, $i=1,\ldots,m$ is to solve the
following optimization problem
\begin{equation}
\left. \begin{array}{cl}
         \displaystyle{}\min_{\alpha_1,\ldots,\alpha_m}  &
                          \displaystyle{} \sum_{i=1}^m \log\left(1+ \frac{\gamma_i}{\alpha_i}\right),\\
         \mbox{subject to } & \displaystyle{}  \sum_{i=1}^m  \alpha_i=1-\alpha,\\
                   &   \alpha_i>0.
         \end{array}
\right.
\label{eqn:MAC:NoisePart}
\end{equation}
This is a convex optimization problem whose solution can be
explicitly found
 by Lagrangian methods~\cite{Convex} as follows (see Appendix~\ref{app:Noise} for details),
\begin{align}
\alpha_i^* = \frac{\sqrt{\gamma_i(\gamma_i+4\mu)}-\gamma_i}{2}, \ i=1,\ldots,m, \nonumber
\end{align}
where $\mu$ satisfies
\begin{equation}
          \frac{1}{2}\sum_{i=1}^m  \left(\sqrt{\gamma_i(\gamma_i+4\mu)}-\gamma_i\right) = 1-\alpha. \label{eqn:MAC:noise:mu}
\end{equation}

Although solving this problem in closed-form is challenging, as the specific value of $\mu$ depends both on $\alpha$
and the relative magnitude of all $\gamma_i$, its upper and lower
bounds can be determined as shown by
Lemma~\ref{lemma:MAC:mu} below. Since the LHS of
\eqref{eqn:MAC:noise:mu} is monotonously increasing with respect
to $\mu$, it is simple to find $\mu$ numerically by
evaluating \eqref{eqn:MAC:noise:mu} within the region defined by
Lemma~\ref{lemma:MAC:mu}.

\begin{lemma}\label{lemma:MAC:mu}
Given $\alpha\in[0,1]$  and $\gamma_i >0$,  $i=1,...,m$, the $\mu$
defined by \eqref{eqn:MAC:noise:mu} is bounded by
\begin{align}\label{eqn:lemma:MAC:mu}
\frac{1-\alpha}{m} + \frac{(1-\alpha)^2}{m} \frac{1}{\sum_i \gamma_i} \leq \mu \leq \frac{1-\alpha}{m} + \frac{(1-\alpha)^2}{m^2}\frac{1}{\min_i \gamma_i},
\end{align}
where both equalities hold  if and only if
$\gamma_1=\ldots=\gamma_m$.
\end{lemma}
\begin{IEEEproof}
See Appendix~\ref{Proof:lemma:MAC:mu}.
\end{IEEEproof}

From Lemma~\ref{lemma:MAC:Rs} below, we can see that the freedom of
adjusting $\alpha\in[0,1]$ in the optimized noise partition
\eqref{eqn:MAC:NoisePart} can not improve the sum rate constraint
$R_{MAC}$. This is intuitive as $R_{MAC}$ is achievable when all
the source nodes can fully cooperate.

\begin{lemma}\label{lemma:MAC:Rs}
Given
$\gamma_1, \ldots, \gamma_m>0$, for any $\alpha\in[0,1]$, we have
\begin{align}
\min\{R_s (\alpha), \sum_{i=1}^m R_i(\alpha)\} \geq R_{MAC},
\end{align}
with equality when $\alpha=1$.
\end{lemma}
\begin{IEEEproof}
See Appendix~\ref{proof:lemma:MAC:Rs}.
\end{IEEEproof}

\subsection{Upper Bounding Models for Independent BCs with $m$ Receivers}\label{sec:Eqv:BC:upper}

The upper bounding model for independent BCs with $m$ receivers
can be generalized straightforwardly from~\cite{FlavioITW}  as
follows
\begin{align}
\mbC_{u,BC,1}= & ( R_{BC},\log(|\mathcal{Y}_1|), \ldots, \log(|\mathcal{Y}_m|)), \label{eqn:BC:upp:shot1}\\
\mbC_{u,BC,2}= & (\log(|\mathcal{X}|), R_1, \ldots, R_m), \label{eqn:BC:upp:shot2}
\end{align}
where
\begin{align}
R_{BC} & \triangleq  \max_{p(x)} I(X; Y_1,\ldots,Y_m),\label{eqn:BC:upp:shot-sum}\\
R_i  & \triangleq  \max_{p(x)} I(X; Y_i), \ i=1,\ldots,m. \label{eqn:BC:upp:shot-i}
\end{align}
Note that $\mbC_{u, BC, 1}$ is a valid upper bound for any
channel transition function $p(y_1,\ldots,y_m|x)$ whereas
$\mbC_{u, BC, 2}$ is only valid for BC with independent
noise components at receivers, i.e., when the transition
probability can be factorized as
\[p(y_1,\ldots,y_m|x)=\prod^m_{i=1}p(y_i|x).\]

\subsubsection{New Upper Bounding Models for $m$-user BCs}\label{sec:Eqv:BC-new}
We construct a new upper bounding model by combining the point-to-point
channel emulation technique developed in~\cite{NetEquiv} with the
\emph{Covering Lemma}, \emph{Conditional Typicality
Lemma}, and the \emph{Joint Typicality
Lemma}~\cite{NetworkIT}. Let $[l_1, l_2, \ldots, l_m]$ denote a permutation of the $m$ receivers and
$[Y_{1}, Y_{2}, \ldots, Y_{m}]$ be their
corresponding channel outputs, whose dependence is characterized by the channel transition function $p(y_{1},\ldots,y_{m}|x)$.
The new bounding model is represented by
\begin{align}
\mbC_{u, BC, new} & = (R_{s}, R_{l_1}, R_{l_2}, \ldots, R_{l_m}),\label{eqn:BC:upp:New}\\
 R_{l_k} & =  \max_{p(x)} I(X;Y_1,\ldots,Y_k), \ k{=}1,\ldots,m, \nonumber\\
 R_s  & = R_{l_m} = \max_{p(x)} I(X;Y_1,\ldots,Y_m), \nonumber
\end{align}
where $R_s$ is the sum rate constraint from the transmitter to the auxiliary node $n_I$, and $R_{l_k}$ is the rate constraint from $n_I$ to receiver $l_k$.

\begin{remark}
Compared to the upper bounding model  $\mbC_{u, BC, 1}$ specified in \eqref{eqn:BC:upp:shot1}, the new model $\mbC_{u, BC, new}$
maintains the tight sum rate constraint $R_{BC}$ as specified in \eqref{eqn:BC:upp:shot-sum} and meanwhile improves all the individual
rate constraints.
\end{remark}

Below we show step-by-step how to construct the new bounding model given in \eqref{eqn:BC:upp:New}.
To simplify notation, for $k{=}1,\ldots,m$, let $Y_{[1:k]}$ represent the sequence of $k$ random variables $\{Y_1,Y_2,\ldots,Y_k\}$ and $w_{[1:k]}$ be the sequence of $k$ integers $(w_1,w_2,\ldots,w_k)$.
The channel emulation is done over a stacked network with $N$ replicas of
the original BC.

\textbf{Step I}: Fix a channel input distribution $p_X(x)$.
As defined in~\cite{NetEquiv}, let  $\hat{A}^{(N)}_{\epsilon}(X)$ be a subset of
the classical typical\footnote{In this subsection, the typical set $T^{(N)}_{\epsilon}$ is defined based on the strong typicality to support the \emph{Conditional Typicality Lemma} and the \emph{Joint Typicality Lemma}.} set $T^{(N)}_{\epsilon}(X) $\cite{NetworkIT} such that
for any $x^N\in \hat{A}^{(N)}_{\epsilon}(X)$ as the input to the
BC, the probability that the corresponding output sequences $\{y_1^N,y_2^N,\ldots,y_m^N\}$
are not jointly typical with $x^N$ is smaller than a predefined threshold.
Furthermore, let
$p_{Y_{[1:k]}}(y_1,\ldots,y_k)$, $k{=}1,\ldots,m$,  be marginal distributions
obtained from $p(x,y_1,\ldots,y_m)=p(y_1,\ldots,y_m|x)p_X(x)$, and define a series  of conditional distributions as follows
\begin{align} \label{eqn:BC:upp:pyy}
p(y_{k+1}|y_{[1:k]}) \triangleq \left\{
\begin{array}{cc}
0, & \mbox{ if } p_{Y_{[1:k]}}(y_{[1:k]})=0,\\
\frac{p_{Y_{[1:k+1]}}(y_{[1:k{+}1]})}{p_{Y_{[1:k]}}(y_{[1:k]})}, & \mbox{otherwise}.
\end{array}
\right.
\end{align}

\textbf{Step II}: Generate independently at random $2^{NR'_{1}}$ sequences
$\{y_1^N(w_1): w_1{=}1,\ldots,2^{N R'_{1}}\}$, each according to $\prod_i
p_{Y_1}(y_{1,i})$. For any sequence $x^N{\in} \hat{A}^{(N)}_{\epsilon}(X)$, by the
\emph{Covering Lemma}~\cite{NetworkIT},
\begin{align}\label{eqn:BC:u1a}
\lim_{N \to \infty} Pr\left(\exists w_1{\in}[1:2^{N R'_1}], \mbox{s.t. }(x^N, y_1^N(w_1)){\in} T^{(N)}_{\epsilon_1}\right) = 1,
\end{align}
if $R'_1>I(X;Y_1) + \delta_1(\epsilon_1)$ for some
$\epsilon_1>\epsilon>0$ and $\delta_1(\epsilon_1)>0$ that goes to
zero as $\epsilon_1\to 0$. Following the channel emulation
argument~\cite{NetEquiv}, we define a mapping function
$\alpha_1(x^N)$ as
\begin{align}
\alpha_1(x^N) = \left\{
\begin{array}{lc}
w_1, & \mbox{ if } \exists w_1 \mbox{ s.t. } (x^N, y_1^N(w_1))\in T^{(N)}_{\epsilon_1},\\
1, & \mbox{ otherwise}.
\end{array}
\right.
\end{align}
If  more than one sequence are jointly typical with
$x^N$,  $\alpha_1(x^N)$ chooses one of them uniformly at
random.

\textbf{Step III}: For each sequence $y_1^N(w_1)$, generate
independently $2^{NR'_{2}}$  sequences $\{y_2^N(w_1,w_2):
w_2{=}1,\ldots,2^{N R'_{2}}\}$, each according to $\prod_i
p(y_{2,i}|y_{1,i}(w_1))$, where $p(y_2|y_1)$ is defined in \eqref{eqn:BC:upp:pyy}.
Given $w_1=\alpha_1(x^N)$, according to the \emph{Conditional Typicality
Lemma}~\cite{NetworkIT}, we have
\begin{align}
Pr\left((x^N, y_1^N(w_1))\in
T^{(N)}_{\epsilon_1}(XY_1)\right)\to 1 \mbox{ as } N\to\infty,
\end{align}
and according to the
\emph{Joint Typicality Lemma}~\cite{NetworkIT}, for all $w_2\in[1:2^{N R'_{2}}]$,
\begin{align}\label{eqn:BC:u1b}
Pr\big((x^N, y_1^N(w_1), y_2^N(w_1,w_2))\in & T^{(N)}_{\epsilon_2}(XY_1Y_2) \big) \\
 & \geq 2^{-N(I(X;Y_2|Y_1) + \delta_2(\epsilon_2))},\nonumber
\end{align}
for $\epsilon_2>\epsilon_1$ and some $\delta_2(\epsilon_2)>0$ that
goes to zeros as $\epsilon_2\to 0$.
Since sequences $\{y_2^N(w_1,\cdot)\}$ are i.i.d given $w_1$, we have
\begin{align}
& Pr\left(\forall w_2\in[1:2^{N R'_{2}}],  (x^N, y_1^N(w_1), y_2^N(w_1,w_2))\notin T^{(N)}_{\epsilon_2}\right) \nonumber\\
& =\left( Pr\left((x^N, y_1^N(w_1), y_2^N(w_1,w_2))\notin T^{(N)}_{\epsilon_2} \right) \right)^{2^{N R'_2}} \nonumber\\
 & \leq \left( 1- 2^{-N(I(X;Y_2|Y_1) + \delta_2(\epsilon_2))}  \right)^{2^{N R'_2}}\label{eqn:BC:u1c}\\
& \leq exp\left(-2^{N R'_2}\cdot 2^{-N(I(X;Y_2|Y_1) + \delta_2(\epsilon_2))} \right)\nonumber\\
& =exp\left(- 2^{N(R'_2 - I(X;Y_2|Y_1) - \delta_2(\epsilon_2))} \right),\nonumber
\end{align}
where the first inequality comes from \eqref{eqn:BC:u1b} and the second inequality comes from the fact that $(1-x)^n\leq exp(-nx), \forall x\in[0,1]$.
Therefore if $R'_{2}>I(X;Y_2|Y_1) + \delta_2(\epsilon_2)$, the probability that none of the sequences in $\{y_2^N(w_1,w_2): w_2\in[1:2^{N R'_{2}}]\}$  is joint typical with $(x^N, y_1^N(w_1))$ tends to  $0$ as $ N \to \infty$.
We now define a mapping function $\alpha_2(x^N, w_1)$ as follows
\begin{align}
\alpha_2(x^N, w_1) {=} \left\{
\begin{array}{lc}
w_2, & \mbox{ if } (x^N, y_1^N(w_1), y_2^N(w_1,w_2))\in T^{(N)}_{\epsilon_2},\\
1, & \mbox{ otherwise}.
\end{array}
\right.
\end{align}
If there is more than one candidate that satisfies the joint typicality
condition,   $\alpha_2(\cdot)$ chooses one of them uniformly at
random.

\textbf{Step IV}: For $k=3,\ldots,m$, we treat the set of sequences $\{y_1^N(w_1),\ldots,y_{k-1}^N(w_{[1:k-1]})\}$ together as one unit and repeat Step III, which generates the corresponding sequences $\{y_k^N(w_{[1:k-1]}, w_k):
w_k{=}1,\ldots,2^{N R'_{k}}\}$, the mapping function $\alpha_k(x^N, w_{[1:k-1]})$, and the rate constraint
\begin{align}\label{eqn:BC:upp:Rk}
R'_{k}>I(X;Y_k|Y_{[1:k-1]}) + \delta_k(\epsilon_k),
\end{align}
where $\epsilon_k>\epsilon_{k-1}$, and $\delta_k(\epsilon_k)>0$ that
goes to zeros as $\epsilon_k\to 0$.

\textbf{Step V}:
Define a channel emulation codebook
\begin{align}
\{ y_k^N(w_1,\ldots,w_k): k{=}1,\ldots,m, w_k{=}1,\ldots,2^{R'_k}\},
\end{align}
and the associated encoding function $\alpha(x^N){=}[\alpha_1(\cdot),\ldots,\alpha_m(\cdot)]$ and the decoding function $\alpha^{-1}_k(w_{[1:k]})$ for receiver $l_k$, $k{=}1,\ldots,m$.
For any input $x^N$, $\alpha(x^N)$ generates a sequence $(w_1, w_2,\ldots, w_m)$ of $N\sum_{i=1}^m R'_i$ bits that are transmitted from the transmitter of the BC to the auxiliary node $n_I$, which then forwards them to receiver $l_k$. At receiver $l_k$, the decoding  function $\alpha^{-1}_k(w_{[1:k]})$ selects a sequence from the codebook $\{y_k^N\}$ based on the received information bits, i.e.,
\[\alpha^{-1}_k(w_1,\ldots,w_k) = y_k^N(w_1,\ldots,w_k).\]

Note that the rate constraints in \eqref{eqn:BC:upp:Rk} should be satisfied for $k{=}1,\ldots,m$, and for
all $p_X(x)$. Let  $N\to \infty$ and $\epsilon_m\to 0$\footnote{Since $0<\epsilon<\epsilon_1<\ldots < \epsilon_m$, letting
$\epsilon_m\to 0$ implies that all of them go to zero.},
we can specify all the rate constraints in   \eqref{eqn:BC:upp:New} as follows
\begin{align}
R_{l_k} & = \sum_{i=1}^k R'_i = \max_{p(x)} I(X;Y_1,\ldots,Y_k), \ k{=}1,\ldots,m, \label{eqn:BC:upp:New2} \\
R_s &= R_{l_m} = \max_{p(x)} I(X;Y_1,\ldots,Y_m).\label{eqn:BC:upp:New3}
\end{align}
The second equality in \eqref{eqn:BC:upp:New2} comes from the fact that
\begin{align}
I(X;Y_{[1:k]}) {=} I(X;Y_1) {+} I(X;Y_2|Y_1) {+} \ldots {+} I(X;Y_k|Y_{[1:k{-}1]}),
\end{align}
and the first equality in \eqref{eqn:BC:upp:New3} comes from our emulator design  as specified in Step V.

\begin{remark}
There are in total $m!$ different permutations of
$l_1,\ldots,l_m$, each leading to a different upper bounding
model following our construction method. For each of these upper bounding models, the sum rate constraint and
one of the individual rate constraints are tight. Depending on the needs, we can select a specific permutation to design the upper bounding model.
\end{remark}

\begin{remark}
For BC with $m=2$ receivers, the proposed upper bounding model has two different layouts
$\mbC_{u, BC, a}=(R_{BC}, R_1, R_{BC})$ and $\mbC_{u, BC, b}=(R_{BC}, R_{BC}, R_2)$, where the latter turns out
 to be equivalent to the upper bounding model
developed in~\cite[Theorem~5]{NetEquiv2}. This is not surprising
as the channel emulation codebook $\{y_1^N, y_2^N\}$ used in our
construction is generated in the same way as
in~\cite[Theorem~5]{NetEquiv2}: superposition encoding.
Note that the proof in~\cite[Theorem~5]{NetEquiv2}, restricted for BC with $m{=}2$
receivers, provides explicit error analysis. In contrast, our construction is valid for  $m{\geq}2$ but only claims that the error probability can be made arbitrarily small with the help of the \emph{Covering Lemma}, the \emph{Conditional Typicality Lemma}, and the \emph{Joint Typicality Lemma}.
\end{remark}

\subsubsection{Extension to $m$-user BCs with Continuous Alphabets}\label{sec:Eqv:BC-gauss}

The error analysis in our construction of new upper bounding models for $m$-user BCs in Sec.~\ref{sec:Eqv:BC-new} relies on the validity of
the \emph{Covering Lemma}, the \emph{Conditional Typicality Lemma}, and the \emph{Joint Typicality Lemma}, which
 hold for discrete-alphabet channels under strong/robust typicality notions.
 There are several possible approaches to extend our results  in Sec.~\ref{sec:Eqv:BC-new} to  continuous alphabets.

  One way is to apply the standard discretization procedure~\cite[Chp.~3.4.1]{NetworkIT} to continuous alphabets\footnote{As described in~\cite[Chp.~3.4.1]{NetworkIT}, the transition function of the continuous-alphabet channel should be ``well-behaved'' to facilitate the discretization and quantization procedure.} and then apply the results derived based on discrete alphabets. Such process has been demonstrated in \cite[Chp.~3.4.1]{NetworkIT} to prove the achievability of AWGN channel capacity, and in \cite{Jalali11} to extend separation results from DMC to AWGN channels with a per-symbol average power constraint.

Another way is to use generalized typicality definitions such that the \emph{Covering Lemma} (which essentially depends on the \emph{Conditional Typicality Lemma}) and the \emph{Joint Typicality Lemma} can be extended to continuous alphabets.
 For example, joint typicality properties associated with strong typicality have been extended to countably infinite alphabets in \cite{UniType10} by a notion of unified typicality.
 The \emph{Generalized Markov Lemma} has been extended in \cite{Oohama98} to Gaussian sources with a modified notion of typicality and by the fact that the asymptotic equipartition property (AEP) holds for Gaussian memoryless sources.
 Other attempts in this direction can be found, for example, in \cite{Jeon14} where the \emph{covering lemma} and \emph{packing lemma} are extended to continuous alphabets (memoryless source) and in \cite{softCover14} where the likelihood encoder \cite{Cuff13} with the soft-covering lemma \cite{Wyner75, Cuff13} is applied to the lossy source compression with continuous alphabets.

We may also construct our new upper bounding models in Sec.~\ref{sec:Eqv:BC-new} by iteratively applying the method
developed in~\cite[Theorem~5]{NetEquiv2}: We first only focus on two channel outputs $Y_1, Y_2$, and contract a valid upper bound for this 2-user BC $\mX\to \mY_1\times\mY_2$ as in~\cite[Theorem~5]{NetEquiv2}; then we treat $[Y_1, Y_2] \in \mY_1\times\mY_2$  as a single compound receiver and group it with $Y_3$ to formulate a new $2$-user BC, whose upper bound can also be constructed as in~\cite[Theorem~5]{NetEquiv2}; we can repeat this process until there is no more receivers to be included. However, the associated error analysis can be very involved.

%

\subsection{Lower Bounding Models for Independent MACs and BCs}\label{sec:noncoup:lower}

Our lower bounding models for MACs/BCs are constructed directly
based on some operating points within the achievable rate region
assuming no transmitter/receiver cooperation. However, the
structure of the  lower bounding models are quite
different for MACs and BCs. The main difference between MACs and
BCs is the encoding process. When there is no transmitter/receiver
cooperation, distributed encoding is performed in MACs while
centralized encoding is done in BCs. As a consequence, in MAC
setups, only one rate constraint is needed for each point-to-point
bit pipe to fully describe any operation point within the rate
region.
In BCs, each of the private messages
 dedicated for one specific receiver may also be decoded by other
 receivers.
 Such ``overheard'' messages (the common messages) should be
 reflected in the rate region, which requires the usage of point-to-points bit pipes (hyper-arc) in the lower bounding model.

\subsubsection{MACs with $m$ Transmitters}\label{sec:Eqv:MAC:lower}

The lower bounding models can be constructed by choosing an
operating point in the capacity region of the MAC assuming
independent sources. We can choose any point in the capacity region that can be achieved by
using independent codebooks at transmitters and successive
interference cancellation decoding at the receiver. For Gaussian MAC with
$m$ transmitters, each with received SNR $\gamma_i$,
$i=1,\ldots,m$, the following sum rate is achievable
\begin{equation}\label{eqn:MAC:low:sum}
R_{l,s} =  \frac{1}{2}\log\left(1 + \sum_{i=1}^m \gamma_i \right).
\end{equation}

\subsubsection{BCs with $m$ Receivers}\label{sec:Eqv:BClower}

\begin{table}[t]
\centering \caption{All possible rate constraints in the lower bounding model
for broadcast channels with $m$ receivers.}
\label{tab:BC_R}
\begin{tabular}{|c|c|c|c|c|c|}
\hline
 $R_i, i=$        &     1         &   2           &       3       & $\cdots$ & $2^m-1$\\
 \hline
 $i:\{0,1\}^m$ & $0\cdots 001$ & $0\cdots 010$ & $0\cdots 011$ & $\cdots$ & $1\cdots111$ \\
 \hline
 $\{\mD_n\}, n=$  & $1$          &   $2$     & $2, 1$ &  $\cdots$   & $m,\cdots,2,1$  \\
 \hline
\end{tabular}
\end{table}

For BCs, each of the private messages dedicated for one specific
receiver may also be decoded by other receivers, and such
overheard messages can be useful when the BCs are part of a larger
network. To model such message overhearing, we need to introduce point-to-points bit pipes (i.e., hyper-arcs) to represent multicast
rate constraints.
For BCs with $m$ receivers $\{\mD_n$, $n=1,\ldots,m\}$, there are in total
$(2^m{-}1)$ subsets of receivers, each corresponding to a unique rate constraint.
As illustrated in
Table~\ref{tab:BC_R}, we denote $R_i$ as the rate constraint corresponding to successful decoding at receivers
indicated by the locations of `$1$' in the
length-$m$ binary expression of the index $i$. For example, $R_3$
is the constraint for the multicast rate to receivers $\mD_2$ and
$\mD_1$, and $R_{2^m-1}$ is the constraint for multicast rate to
all receivers. Depending on the channel quality, we represent the lower bounding mode by a vector\footnote{For each $R_i$
in $\bR$ we also need to store its index $i$ to specify the receiving subset.} $\bR$ which contains
one sum rate constraint (denoted by $R_0$) and  up to $m$ constraints\footnote{For statistically degraded
$m$-receiver BCs (e.g., Gaussian BCs), $m$ constraints are sufficient by creating a
 physically degraded channel via proper coding schemes, and the rate loss will
 vanish in low SNR regime~\cite{Fawaz11}. For non-degraded channels,
we only focus on the first $m$ most significant non-zero constraints.}
 from Table~\ref{tab:BC_R}. We illustrate this by an example of Gaussian BCs with $m$ receivers.

\textbf{Example: Gaussian BCs with $m$ Receivers}
Let $\gamma_i$ be the received SNR at
receiver $\mD_i$, $i=1,\ldots,m$. Without loss of
generality, assuming $\gamma_1\leq \gamma_2\leq ...\leq \gamma_m$,
we divided the total information into $m$ distinct messages
$\{W_i, i=1,\ldots,m\}$. By superposition coding of $W_i$ with
power allocation parameters $\beta_i\in[0,1]$,  $\sum_{i=1}^m
\beta_i{=}1$, at the transmitter, and successive interference
cancellation at each receiver\footnote{Alternatively, one can
encode $W_1$ to $W_m$ successively by dirty paper
coding~\cite{Costa} and use a maximum likelihood decoder at each
receiver.}, successful decoding of $W_i$ can be realized at a set
of receivers $\{\mD_n, n=i,\ldots,m\}$ with multicast/unicast rate
\begin{align}\label{eqn:Eqv:BC-Ri}
R_{2^m-2^{i-1}} = \frac{1}{2}\log\left(1 + \frac{\beta_i\gamma_i}{1+\gamma_i\sum_{j=i+1}^m \beta_j}\right).
\end{align}
For example, successful decoding of $W_1$ can be realized at all
receivers with a multicast rate of $R_{2^m -1}$, and successful
decoding of $W_m$ can only be realized at receiver $\mD_m$ with a
unicast rate of
 $R_{2^{m-1}}$.
 The resulting rate vector is therefore
\begin{align}\label{equ:BC:gaus}
 \bR = [R_0, R_{2^m-2^{i-1}}: \ i=1,\ldots,m],
 \end{align}
where the sum-rate constraint $R_0$  is
\begin{align}
R_0 &  = \sum_{i=1}^m R_{2^m-2^{i-1}}
       {=} \sum_{i=1}^m  \frac{1}{2} \log\left(\frac{1+\gamma_i \sum_{j=i}^m
\beta_j}{1{+}\gamma_i \sum_{j=i+1}^m \beta_j}\right)\\
    & = \frac{1}{2}\log(1+\gamma_1) + \frac{1}{2}\sum_{i=2}^m \log\left(\frac{1+\gamma_i \sum_{j=i}^m
\beta_j}{1{+}\gamma_{i-1}\sum_{j=i}^m \beta_j}\right).
\end{align}
The last equality comes from the fact that
\[\sum_{i=1}^m \beta_i=1.\]
 Since $\gamma_{i-1}\leq \gamma_{i}$, the function
$\frac{1+x\gamma_i}{1+x\gamma_{i-1}}$ is monotonically increasing
on $x\in[0,1]$, with its maximum $\frac{1+ \gamma_i}{1+
\gamma_{i-1}}$ achieved when $x=1$. It is simple to show
that
\begin{align}\label{eqn:BC:sum:R0}
R_{0}\leq\frac{1}{2}\log(1+\gamma_m),
\end{align}
where the equality is achieved when $\beta_m=1$ (i.e, $\beta_i=0$
for all $i\neq m$).

\begin{remark}
Note that power allocation at the transmitter of a BC allows
elimination of weakest receivers. For example,  by setting
$\beta_1=\beta_2=0$ in \eqref{eqn:Eqv:BC-Ri} the weakest two
receivers $\mD_1$ and $\mD_2$ will have nothing to decode and
hence be removed from the set of destinations.
\end{remark}

\subsection{Gaps between Upper and Lower Bounding Models for Gaussian MACs and BCs}

For Gaussian MACs with $m$ transmitters, the sum rate is upper
bounded by $R_{MAC}$ given by \eqref{eqn:MAC:upp:sum}, and lower
bounded by $R_{l,s}$ given by \eqref{eqn:MAC:low:sum}. The gap
between the upper and the lower bounds on sum rate, measured in
bits per channel use, is therefore bounded by
\begin{align}
\Delta_{MAC}  & = R_{MAC}-R_{l,s}=  \frac{1}{2}\log\left(\frac{1 + (\sum_{i=1}^m \sqrt{\gamma_i})^2}{1 + \sum_{i=1}^m \gamma_i}\right) \nonumber\\
   & \leq \frac{1}{2}\log\left(\frac{1 + m \sum_{i=1}^m \gamma_i}{1 + \sum_{i=1}^m \gamma_i}\right)
          < \frac{1}{2}\log(m),\label{eqn:Eqv:Dmac}
\end{align}
where the first inequality comes from Jensen's inequality based on
the convexity of the function $f(x)=x^2$.  Hence, for Gaussian
MACs with transmitters in isolation, feedback and transmitter
cooperation can increase the sum capacity by at most
$\frac{1}{2}\log(m)$ bits per channel use.

For Gaussian BCs with $m$ receivers, the sum rate is lower bounded
by $R_0$ given by \eqref{eqn:BC:sum:R0} and upper bounded by
\begin{align}
R_{BC}=\frac{1}{2}\log \left(1+ \sum_{i=1}^m \gamma_i\right).
\end{align}
Note that $R_{BC}$ can be achieved only when full cooperation
among all receivers is possible. The gap between the upper and the
lower bounds on the sum rate is therefore
\begin{align}
\Delta_{BC} & =R_{BC}-R_{0}=\frac{1}{2}\log\left(\frac{1+\sum_{i=1}^{m}\gamma_i}{1+\gamma_m}\right) \nonumber\\
& \leq \frac{1}{2}\log\left(\frac{1 + m\gamma_m}{1+\gamma_m}\right)<\frac{1}{2}\log(m),\label{eqn:Eqv:Dbc}
\end{align}
where the first inequality comes from the assumption $\gamma_i\leq
\gamma_m$ for all $i$. Hence, for $m$-receiver Gaussian BCs with
all receivers in isolation, feedback and receiver cooperation can
increase the sum capacity by at most $\frac{1}{2}\log(m)$ bits per
channel use.

 The gap between upper and lower bounding models becomes considerably smaller at low SNR or
when the SNR for each link diverges.  For example, with
$\gamma_1=1, \gamma_2=2, \gamma_3=100$ (e.g, 0, 3, 20dB,
respectively), the gaps (measured in bits per channel use) are
\begin{align*}
\Delta_{MAC} & = \frac{1}{2}\log\left(\frac{1 + (1+\sqrt{2}+10)^2}{1 + 1+2+100}\right) \approx 0.29,\\
\Delta_{BC} & = \frac{1}{2}\log\left(\frac{1+1+2+100}{1+100}\right) \approx 0.02,
\end{align*}
which are much smaller than $\frac{1}{2}\log(3)\approx
0.79$.

\section{Bounding Models for Coupled Networks}\label{sec:Eqv:coup}

Capacity bounding models for MACs and BCs developed
in~\cite{NetEquiv2,Effros1,Effros2, FlavioITW} and  extensions
presented in Sec.~\ref{sec:Eqv:noncoup} are all designed for
networks with non-coupled MACs/BCs. In wireless networks, however,
a signal dedicated for one receiver may also be overheard by its
neighbors, owing to the broadcast nature of wireless transmission.
A transmit signal can be designed for multiple destinations (as in
BCs) and the received signal may consist  of signals from several
source nodes (as in MACs) and thus interfere with each other.
Although such dependence among coupled BCs and MACs has been
partially treated in~\cite{NetEquiv2} by grouping coupled
transmitter-receiver pairs together as ICs, whose bounding models
require up to $m(2^m{-}1)$ bit pipes for IC with $m$
transmitter-receiver pairs, the whole family of multi-hop non-layered channels
(e.g., relay channels) remains untreated.

Inspired by the idea of separate sum and individual rate
constraints~\cite{FlavioITW}, we choose to incorporate dependent
transmitter-receiver pairs into coupled BCs and MACs. With the
help of a new channel decoupling method introduced below, we can decompose a coupled sub-network into  decoupled BCs and MACs. The noisy channel that connects a pair of decoupled BC and MAC is split into two parts by channel decoupling, and their dependence will be taken into account when constructing  bounding models for the BC and the MAC.

\subsection{Channel Decoupling}\label{sec:coupled:decoup}

Given a memoryless network
\[\mbN_T = (\prod_{v}\prod_{i} \mX_{v,i},  p(\by|\bx), \prod_{v} \prod_{j} \mY_{v,j}),\]
where $\mX_{v,i}$ is the input alphabet of the $i$-th outgoing channel and $\mY_{v,j}$ is the output alphabet of the $j$-th incoming channel associated with node $v$, all the channel input-output dependence can be fully characterized by the transition function  $p(\by|\bx)$.  Based on the dependence of input/output alphabets, we rewrite the transition function
into product format and partition the network $\mbN_T$ into \emph{independent} point-to-point channels
$\mbN_{i}$, MACs $\mbN_{S_j\to j}$, BCs $\mbN_{k\to D_k}$, and sub-networks $\mbN_{S_l\to D_l}$.
A sub-network $\mbN_{S_l\to D_l}$ can be further decomposed into independent channels and smaller sub-networks as long as there exist non-trivial complementary subsets $(S,S^c)$ of $S_l$ and $(D,D^c)$ of $D_l$ such that
\begin{align}
p(\by_{D_l}|\bx_{S_l}) = p(\by_{D}|\bx_{S}) p(\by_{D^c}|\bx_{S^c}).
\end{align}
We repeat this partition process until we can't decompose the sub-networks any further.

 To highlight all independent components within the network, we rewrite the network expression as
\begin{align}\label{eqn:network:decoupling}
\mbN_T = \prod_{i} \mbN_{i} \times \prod_{j} \mbN_{S_j\to j} \times \prod_{k} \mbN_{k\to D_k}  \times \prod_{l} \mbN_{S_l\to D_l},
\end{align}
where $\prod$ and $\times$ indicate channel independence as defined in Definition~\ref{dfn:indchannel}.
To construct capacity bounding models for $\mbN_T$, we replace each independent point-to-point channel by a bit pipe with rate equal its channel capacity as prescribed by~\cite{NetEquiv}. For all the independent MACs and BCs, we replace them by their corresponding upper and lower bounding models developed in Sec.~\ref{sec:Eqv:noncoup}. The independent sub-networks need some special treatment as we will discuss below.

\begin{figure}[t]
\centering
\psfrag{xi2}[][]{$n_I$}\psfrag{yj2}[][]{$n_J$}\psfrag{nij}[][]{}
\psfrag{x1}[][]{$\mX_1$}\psfrag{xi}[][]{$\mX_i$}\psfrag{xp}[][]{$\mX_n$}
\psfrag{y1}[][]{$\mY_1$}\psfrag{yj}[][]{$\mY_j$}\psfrag{yq}[][]{$\mY_m$}
\psfrag{di}[][]{}
\psfrag{lj}[][]{$\mZ_{i,j}, \tilde{\mY}_{i,j}$}\psfrag{l1}[][]{$\mZ_{i,1}, \tilde{\mY}_{i,1}$}\psfrag{lq}[][]{$\mZ_{i,m}, \tilde{\mY}_{i,m}$}
\psfrag{ls}[][]{}\psfrag{d1}[][]{}\psfrag{ds}[][]{}\psfrag{dp}[][]{}
\psfrag{eq}{$ \Rightarrow$}
  \includegraphics[width=8.6cm]{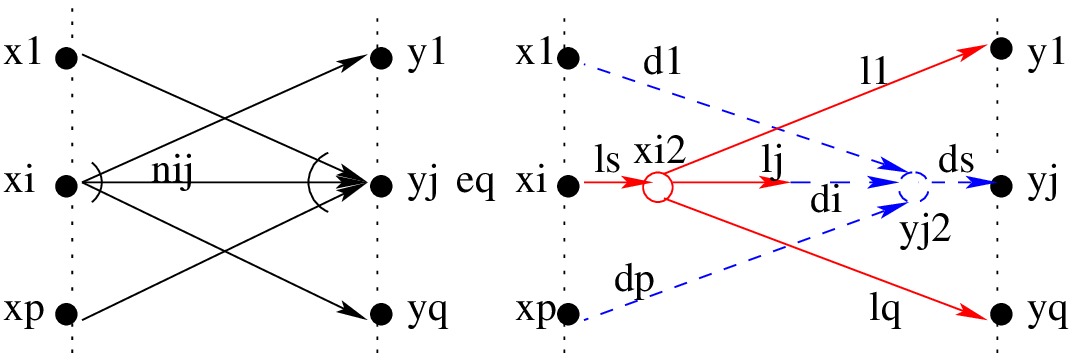}
    \caption{ An independent  sub-network $\mbN=(\prod_{l=1}^n \mX_l, p(\by|\bx), \prod_{k{=}1}^m \mY_k)$ in which a BC with input alphabet $\mX_i$ and a MAC with output alphabet $\mY_j$ are coupled by the noisy connection
$\mX_i\to\mY_j$. After channel decoupling, the decoupled MAC is characterized by $p(y_j|\bx)$ given in \eqref{eqn:coupled:MAC:pdf},  and the decoupled BC is characterize by $p(\bz_i|x_i)$ given in \eqref{eqn:coupled:BC:pdf}
for upper bounding models and by $\tilde{p}_L(\tilde{\by}_i|x_i)$ given in \eqref{eqn:coupled:BC:pdfL} for lower bounding models. The auxiliary random variables $Z_{i,j}$ are introduced for upper bounding models and $\tilde{Y}_{i,j}$ for lower bounding models.}
    \label{fig:Neq:upper}
\end{figure}

For an independent sub-network $\mbN=(\prod_{l=1}^n \mX_l, p(\by|\bx), \prod_{k{=}1}^m \mY_k)$ that contains $n$ transmitting alphabets and $m$ receiving alphabets, as shown in the left part of Fig.~\ref{fig:Neq:upper},
the BC with input alphabet $\mX_i$ and the MAC with output alphabet $\mY_j$ are coupled due to the noisy connection
$\mX_i\to\mY_j$.
Note that $p(\by|\bx)$ does not necessarily indicate a layered structure of the sub-network; rather it is only an abstraction of the input-output dependence of the network (cf. the relay channel with $p(y,y_r|x,x_r)$).
 Our channel decoupling approach that will be presented in Sec.~\ref{sec:coupled:decoup:pdf} has two essential components: the auxiliary node with its associated input/output alphabets, and the new probability functions designed to describe the decoupled MACs and BCs.  When constructing bounding models for the decoupled MACs and BCs, there will be two individual rate constraints on the  bit pipe corresponding to the connection $\mX_i\to\mY_j$: one from  the decoupled BC with input alphabet $\mX_i$ and one from the decoupled MAC with output alphabet $\mY_j$.
  Since both of the two constraints are introduced to  characterize the same individual rate, they should be simultaneously respected when  constructing the noiseless bounding models.
We demonstrate this principle in Sec.~\ref{sec:Eqv:Upper} for the upper bounding models  and in Sec.~\ref{sec:Eqv:Lower} for the lower bounding models.

\subsubsection{Transition Functions for Decoupled MACs/BCs}\label{sec:coupled:decoup:pdf}

As mentioned in Sec.~\ref{sec:Eqv:intro}, in this paper we assume
that the distortion components in a noisy coupled network are
mutually independent, i.e., the transition probability can be
partitioned as
\[p(\by|\bx)=\prod_{j=1}^m p(y_j|\bx),\]
where
\begin{align}\label{eqn:coupled:MAC:pdf}
p(y_j|\bx)  \triangleq \sum_{\by_{/j}} p(\by|\bx),
\end{align}
is the marginal distribution for $Y_j$ given $\bX$. We characterize the decoupled MAC with received alphabet
 $\mY_j$  by the marginal distribution
$p(y_j|\bx)$, which preserves the possibility of source
cooperation (allowing all possible $p(\bx)$ as in $\mbN$).
We can then construct its upper and lower  bounding models
based on $p(y_j|\bx)$ by following the techniques developed in
Sec.~\ref{sec:noncoup:MAC:upper} and in Sec.~\ref{sec:Eqv:MAC:lower},
respectively. However, the lower bounding models will be updated in Sec.~\ref{sec:Eqv:Lower}  to incorporate potential interference.

 Note that by treating each of the decoupled MACs individually, the resulting capacity upper bound for the sub-network can be looser than the bound when we treat the sub-network without decomposition. For any given $p(\bx)$, we have
\begin{align}
I(\bX;\bY)& =h(\bY)-h(\bY|\bX)=h(\bY)-\sum_{j=1}^m h(Y_j|\bX)\\
          & \leq \sum_{j=1}^m h(Y_j)-h(Y_j|\bX) =  \sum_{j=1}^m I(\bX;Y_j),
\end{align}
where the inequality is due to the correlation among $\bY$. This is intuitive as we can always treat $\bX$ as a compound source and the sub-network as a virtual BC: treating each output channel as an independent channel will result in looser rate constraints. However, as we have discussed earlier,  the bounding models for multiple-input multiple-output sub-networks are difficult to obtain even for the simplest $2\times 2$ setup, and there is no systematic approach for constructing upper and lower bounding models as the size of the sub-network grows.

The decoupled BC with transmit alphabet $\mX_i$  can't be
  fully described by its marginal
\begin{align}
p(\by|x_i) \triangleq \sum_{\bx_{/i}} p(\by|\bx) p(\bx_{/i}|x_i),
\end{align}
since $p(\by|x_i)$ is determined not only by   $p(\by|\bx)$ itself, but also by all the
possible distributions of channel inputs $\bX_{/i}$. Therefore
$p(\by|x_i)$ only provides a  description of the average behavior
of the correlation among different channel outputs but erases both
the explicit dependence of $\bY$ on a specific channel input and
the interaction among different channel inputs $\bX$.

 To preserve the structure of the original coupled network and to capture the input-output dependence in the decoupled BCs,  we define a
transition function with transmit  alphabet $\mX_i$ as
\begin{align}\label{eqn:coupled:BC:pdfL}
\tilde{p}_L(\tilde{\by}_i|x_i) \triangleq p(\by|x_i, \bx_{/i}=\emptyset ),
\end{align}
 where $\bx_{/i}=\emptyset$ represents the scenario that there is no input
 signal\footnote{For additive channels this implies that we force $\bx_{/i}=0$
 even if $0\not\in\mX_j$, $j\neq i$. For multiplicative channels, we set $\bx_{/i}=1$.}  except $X_i=x_i$, and
$\tilde{\bY}_i$ are  auxiliary random variables introduced to represent the corresponding channel output. The
 transition function defined in \eqref{eqn:coupled:BC:pdfL} will
 be first used to construct \textbf{lower bounding models} for the
 decoupled BC using the techniques developed in
 Sec.~\ref{sec:Eqv:BClower}, where a specific coding scheme (e.g., superposition with rate and power allocation) is designed for $X_i$.
 The effect of interference from other signals will be incorporated into the lower bounding models in the two-step update
presented in Sec.~\ref{sec:Eqv:Lower}.
 Note that the definition
 \eqref{eqn:coupled:BC:pdfL} accommodates the explicit dependence
 of channel outputs on the input signal $X_i=x_i$ but erases source
 cooperation. It therefore enables  efficient and simple
 characterization of the individual and sum rate constraints
 (optimized over $p(x_i)$) which are otherwise difficult to obtain
 (optimized over $p(\bx)$). Furthermore, it preserves possible
 noise correlation in $\bY$ which will be useful in future work.

 To construct \textbf{upper bounding models} for decoupled BCs, we
 introduce a group of axillary random variables
 \begin{align}
 \bZ& \triangleq \{Z_{i,j}| i{=}1,\dots,n, j{=}1,\dots,m\} \nonumber\\
    & = \{\bZ_{i}| i{=}1,\dots,n\} = \{\bZ^{j}|j{=}1,\dots,m\},\nonumber
 \end{align}
 and a predefined function
 $\bY=g(\bZ)$ such that
 \begin{align}\label{eqn:coupled:BC:pdf}
p(\by|\bx) = \sum_{\bz: \by=g(\bz)} \prod_{i=1}^n p(\bz_i|x_i),
\end{align}
 where $\bz_i=[z_{i,1}, \ldots, z_{i,m}]$ is the corresponding
 ``output'' vector of the input signal $X_i=x_i$ and $\bz^j=[z_{1,j}, \ldots, z_{n,j}]$ is the decomposed output vector from received signal $Y_j=y_j$.  The corresponding
 upper bounding models for the decoupled BC with transition
 probability $p(\bz_i|x_i)$ can be therefore constructed following
 the techniques developed in Sec.~\ref{sec:Eqv:BC:upper}.
 However, as in the case for decoupled MACs, treating each decoupled BC individually in constructing upper bounding models will result
 in a looser capacity upper bound than treat them all together (if possible).
For any given $p(\bx)$, we have
 \begin{align}
 I(\bX;\bY) & \leq I(\bX;\bZ_1,\ldots,\bZ_n) \label{eqn:couple:BC:Ixy1}\\
            & = h(\bZ_1,\ldots,\bZ_n) - h(\bZ_1,\ldots,\bZ_n|\bX)\\
            & = h(\bZ_1,\ldots,\bZ_n) - \sum_i h(\bZ_i|X_i) \label{eqn:couple:BC:Ixy} \\
            & \leq \sum_i h(\bZ_i) - \sum_i h(\bZ_i|X_i) \label{eqn:couple:BC:Ixy2}\\
            &= \sum_i I(X_i;\bZ_i),
 \end{align}
 where \eqref{eqn:couple:BC:Ixy1} is due to $\bY=g(\bZ)$,  \eqref{eqn:couple:BC:Ixy} comes from
 \eqref{eqn:coupled:BC:pdf} and  the fact that $X_j$---$X_i$---$\bZ_i$ forms a
Markov chain for all $i\neq j$, and \eqref{eqn:couple:BC:Ixy2} is due to chain rule and the fact that condition reduces entropy.

\begin{figure}[t]
\centering
\psfrag{x1}[][]{$\mX_1$}\psfrag{x2}[][]{$\mX_2$}\psfrag{x}[][]{$\mX$}
\psfrag{y1}[][]{$\mY_1$}\psfrag{y2}[][]{$\mY_2$}\psfrag{y}[][]{$\mY$}
\psfrag{A}[][]{(A)}\psfrag{B}[][]{(B)}
\psfrag{z1}[][]{$\mZ_1$}\psfrag{z2}[][]{$\mZ_2$}\psfrag{v1}[][]{$\mV_{1}$}\psfrag{v2}[][]{$\mV_{2}$}
\psfrag{z3}[][]{$\mZ_{11}$}\psfrag{z4}[][]{$\mZ_{12}$}\psfrag{z5}[][]{$\mZ_{21}$}\psfrag{z6}[][]{$\mZ_{22}$}
\psfrag{v3}[][]{$\mV_{11}$}\psfrag{v4}[][]{$\mV_{12}$}\psfrag{v5}[][]{$\mV_{21}$}\psfrag{v6}[][]{$\mV_{22}$}
  \includegraphics[width=8.6cm]{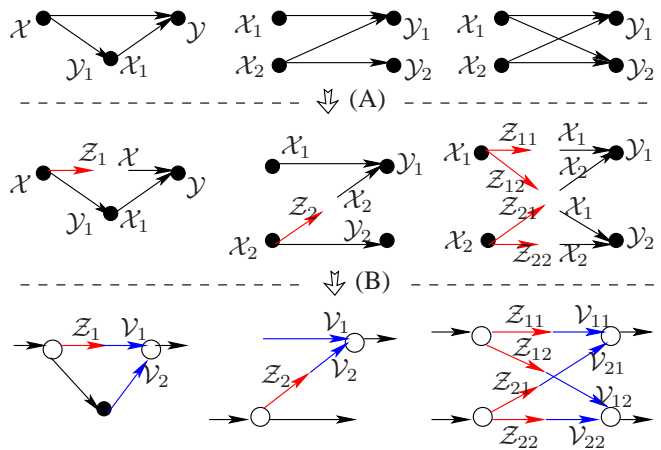}
    \caption{Illustration of channel decoupling (A) and upper bounding process (B) for the relay channel, the two-user Z-channel, and the two-user X-channel. After channel decoupling, the output of the decoupled BCs are changed from $\mY$ to $\mZ$ (highlighted in \textcolor{red}{\textbf{red}}) as described by \eqref{eqn:coupled:BC:pdf} and input to the decoupled MACs inherent the original alphabets as described by \eqref{eqn:coupled:MAC:pdf}.
   To construct upper bounding models for each decoupled two-user MAC, we introduced two auxiliary random variables (highlighted in \textcolor{blue}{\textbf{blue}}) as described by \eqref{eqn:MACnew}--\eqref{eqn:pVX} in Sec.~\ref{sec:noncoup:MAC:upper}.}
    \label{fig:Neq:decoupling}
\end{figure}

In Fig.~\ref{fig:Neq:decoupling} we illustrate the operation of channel decoupling for the three-node relay channel, the two-user Z-channel, and the two-user X-channel. Note that we do not specify the transmission task (e.g., unicast or multicast) for the Z-channel and the X-channel since our bounding models are designed for general transmission schemes rather than tailored/optimized for any specific communication task. For the relay channel, the decoupled MAC is the same as if the source-relay link were not present, and the decoupled BC has an auxiliary random variable $Z_1$ that depends on $X$ but not $X_1$, but meanwhile must reflect the distortion component at the destination when constructing upper bounding models. A different auxiliary random variable $\tilde{Y}_1$ will be introduced here when constructing lower bounding models as described by \eqref{eqn:coupled:BC:pdfL}. It is clear that choosing $Z_1$ and $\tilde{Y}_1$ such that $X$---$Z_1$---$\tilde{Y}_1$ formulates a Markov chain can satisfy all the previous constraint. The decoupling operation for the Z-channel is similar. The X-channel is decoupled into two MACs and two BCs, and for each decoupled BC we introduce auxiliary random variables $Z_{i,j}$ and $\tilde{Y}_{i,j}$ for constructing upper and lower bounding models, respectively.

\subsubsection{Channel Decoupling via Noise Partition for Gaussian Networks}\label{sec:coupled:decoup:Gauss}

For Gaussian coupled networks, the auxiliary
variables $\bZ=\{Z_{i,j}|
 i{=}1,\dots,n, j{=}1,\dots,m\}$ can be determined by the noise partition approach as in
Sec.~\ref{sec:noncoup:MAC:upper}. Let $\bH=[h_{i,j}]_{n\times m}$
be the matrix of channel coefficients such that
\begin{align}
\by=\bH^T\bx + \bw = [\bh_1, \ldots, \bh_n]\bx + \bw,
\end{align}
where $\bh_i$ is the column vector of channel coefficients from
source node $i$ to all receiving nodes, $\bx$ is the transmitting
vector with average power constraint $E[|x_i|^2]\leq P_i$, and
$\bw$ is the vector of noise with zero mean and unit variance. We can
partition the noise components into independent terms such that
\begin{align}\label{eqn:gauss:partition}
\bz_i = \bh_i x_i + \bw_i,
\end{align}
where $\bw_i=[w_{i,1},\ldots,w_{i,m}]^T$ is the vector of
partitioned noise components  w.r.t. the input signal $x_i$ and
the variance of $w_{i,j}$ is denoted by $\alpha_{i,j}>0$. Define
$\gamma_{i,j}=P_i*|h_{i,j}|^2$, the noise partition parameters
$\alpha_{i,j}$ (and hence $z_{i,j}$) can be determined by the
following optimization problem
\begin{equation}
\left. \begin{array}{cl}
         \displaystyle{}\min_{\alpha_{i,j}}  &
                          \displaystyle{} \sum_{i=1}^n \log\left(1+ \sum_{j=1}^{m} \frac{\gamma_{i,j}}{\alpha_{i,j}}\right),\\
         \mbox{subject to } & \displaystyle{}  \sum_{i=1}^n  \alpha_{i,j}=1, \ \forall j\in\{1,\ldots,m\},\\
                   &   \alpha_{i,j}>0.
         \end{array}
\right.
\label{eqn:BC:NoisePart}
\end{equation}
Note that \eqref{eqn:BC:NoisePart} is a convex optimization
problem whose solution can be explicitly found
 by Lagrangian methods~\cite{Convex} as shown in Appendix~\ref{Proof:decouple:gauss}.

\subsection{Upper Bounding Models}\label{sec:Eqv:Upper}%

Given a memoryless coupled noisy network with independent noise, we first apply the channel decoupling method proposed in
Sec.~\ref{sec:coupled:decoup} to decompose it into decoupled MACs and BCs.
The upper bounding models for the decoupled BC with transmit alphabet $\mX_i$ can be constructed by techniques developed in Sec.~\ref{sec:Eqv:BC:upper}, with the transition
 probability $p(\bz_i|x_i)$ given in \eqref{eqn:coupled:BC:pdf},
where the  auxiliary random variables $Z_{i,j}$ that depends only on $X_i$ is introduced by a noise partition approach.
The upper bounding models for the decoupled MAC with receive alphabet $\mY_j$ are constructed
based on the  transition function $p(y_j|\bx)$ given in \eqref{eqn:coupled:MAC:pdf}, and auxiliary random variables $V_{i,j}$ that also only depend on $X_i$ are introduced by \eqref{eqn:MACnew}--\eqref{eqn:pVX} in Sec.~\ref{sec:noncoup:MAC:upper} to reflect the ``contribution'' from input signal $X_i$ to the output signal $Y_j$. We illustrated this upper bounding process in Fig.~\ref{fig:Neq:decoupling} for the relay channel, the two-user Z-channel, and the two-user X-channel. Note that the auxiliary random variables $Z_{i,j}$  defined in \eqref{eqn:coupled:BC:pdf} and $V_{i,j}$ defined in \eqref{eqn:MACnew}--\eqref{eqn:pVX}, though introduced on different stages, depend only on the same $X_i$ and are given by the same ``noise partition'' approach. We can therefore introduce them in such a way  that either $X_i$---$Z_{i,j}$---$V_{i,j}$ or $X_i$---$V_{i,j}$---$Z_{i,j}$ can formulate a Markov chain.
 Then, for each pair of decoupled BC and MAC,the corresponding bit pipe that connects them in the upper bounding model should take the rate constraint that equals the maximum of the two required rates (one from each side).

In the channel decomposition and upper bounding model construction, we have introduced some auxiliary random variables based on the ``noise partition'' approach. Although our approach cannot be applied to \textbf{all channels}, it does apply to at least two types of channels: additive channels with  additive noise, e.g., $Y_j=Z_j + \sum_i H_{ij}X_i$, and multiplicative channels with multiplicative noise, e.g., $Y_j = Z_j\cdot \prod_i H_{ij}X_i$. The former has been demonstrated via the   Gaussian networks in \eqref{eqn:partition:add} and \eqref{eqn:gauss:partition}, and the latter has been illustrated via the two-user binary additive MAC in \eqref{eqn:partition:bin} where the channel can be represented as a binary multiplicative channel $Y=X_1X_2Z$ with alphabets $\{\pm 1\}$. Even though these two channel types cannot be used to represent all channels, they  cover the vast majority of channel models used in publications on wireless communication and networks.

 According to the max-flow
min-cut theorem, the maximum throughput from source to sink can be
no larger than the value of the minimum cut in between. For each
transmission task (unicast or multiple cast), we identify all the
cuts in the resulting upper bounding network (which contains only
noiseless point-to-point connections) and calculate the flows
across each cut. The resulting capacity region is therefore an
outer bound\footnote{As shown in~\cite{AlgebraNC}, the max-flow
min-cut theorem is tight on some noiseless networks, which include
noiseless networks associated with single-source multiple-unicast
transmission, single-source (two-level) multicast transmission,
and multi-source multicast transmission. Therefore the bound we
obtained by the max-flow min-cut theorem might be the capacity
region for the corresponding upper bounding network.} for the
upper bounding network, and hence also an outer bound for the
original coupled noisy network.

\subsection{Lower Bounding Models}\label{sec:Eqv:Lower}

The MACs/BCs lower bounding models presented in Sec.~\ref{sec:noncoup:lower}, see also~\cite{NetEquiv2, Effros1, Effros2, FlavioITW}, are designed for independent MACs and BCs, rather than handling the possible interaction among transmitted/received signals by neighboring nodes in a coupled network. Although such interaction has been considered in~\cite{NetEquiv2} for constructing the lower bounding models for $2\times 2$ ICs, and in~\cite{Kannan14} to approximate the capacity of multiple unicast transmission over bidirectional wireless channels with symmetric fading coefficients, scalable lower bounding models for general transmission tasks over wireless networks are still not available.

After channel decoupling,  the decoupled MAC with receive alphabet $\mY_j$ is
described by $p(y_j|\bx)$ given in \eqref{eqn:coupled:MAC:pdf}, and
the  decoupled BC with transmit alphabet $\mX_i$ is characterized by the transition
 probability $\tilde{p}_L(\tilde{\by}_i|x_i)$ defined in \eqref{eqn:coupled:BC:pdfL}.
 Since the transmit signals $x_i$ from the decoupled BC may contain messages that are not designed for decoding at the decoupled MAC, we
 have to incorporated such interfering signals into account through a two-step update as
presented in Sec.~\ref{sec:coupled:two-step}.
The two-step update is based on the intrinsic feature of successive interference cancellation decoding, where the message currently under decoding suffers interference from non-decoded messages. We start from BCs where a specific coding scheme (e.g., superposition coding with rate splitting and power allocation) is designed for transmitting signal $X_i$ based on $\tilde{p}_L(\tilde{\by}_i|x_i)$,  and then we go to the decoupled MACs where a specific successive decoding order is determined based on $p(y_j|\bx)$ and the encoding strategies that sending $\bx$. Once the decoding order is fixed, we know exactly how much interference one message will suffer from other messages and we can modified the corresponding rate on each bit pipe.

Below we show step-by-step how to construct lower bounding models for decoupled MACs/BCs with a two-step update in Sec.~\ref{sec:coupled:two-step} and demonstrate it via a coupled Gaussian network in Sec.~\ref{sec:two-step:example}.
\subsubsection{Step-by-Step Construction}\label{sec:coupled:two-step}

\textbf{\\ Step I: Network Decomposition}

Apply the channel decoupling method proposed in
Sec.~\ref{sec:coupled:decoup} to decompose the coupled sub-networks
into decoupled BCs and MACs.
\\
\textbf{Step II: Lower Bounding Models for Point-to-Point
Channels, BCs, and Independent MACs}

We replace each independent point-to-point channel with a bit pipe whose rate equals its capacity~\cite{NetEquiv}.
 For each BC and each independent MAC, we
replace them with the corresponding lower bounding models as
described in Sec.~\ref{sec:Eqv:BClower} and in Sec.~\ref{sec:Eqv:MAC:lower}, respectively.
Note that in this step the lower bounding models for decoupled BCs are constructed   assuming no interference.
\\
\textbf{Step III: Lower Bounding Models for Decoupled MACs}

A decoupled MAC has one or more input signals originated from  decoupled BCs.
When constructing lower bounding models in {Step II}, a specific encoding scheme is
adopted by each of the decoupled BCs, whose message might not be  fully decoded at the decoupled MAC.
From the encoding schemes of decoupled BCs, we can identify the signal components that can't be decoded by the decoupled MAC (hence behave as interference).  With such information about the interfering signals, we can construct lower bounding models for the decoupled MAC by integrating the interference effect into the models developed in Sec.~\ref{sec:Eqv:MAC:lower}.
\\
\textbf{Step IV: Rate Adjustment for Decoupled BCs}

The lower bounding models for a decoupled MAC has prescribed a specific decoding strategy, e.g., in which order the messages are decoded. During the successive decoding process,  messages decoded in earlier stages experience higher level of interference. The lower bounding models for the decoupled BCs should be updated to reflect the exact amount of interference encountered during the decoding process. Such update is crucial to ensure that the broadcasting messages from the decoupled BC can be successfully decoded by all intended receivers.

After \textbf{Step IV} we have generate a lower bounding network consists of only noiseless
bit pipes, including point-to-points bit pipes (i.e., hyper-arcs) that carry the same data from one point to multiple
points.  The problem of  finding the optimal scheme to manage the data flows
over such noiseless networks is in general open. However, there
exist many heuristic (and thus suboptimal in general) methods,
see~\cite{NC-multicast} for example, for constructing a valid
inner bound.

\subsubsection{Example: Lower Bounding Models for a Coupled Gaussian $m\times n$ Network}\label{sec:two-step:example}

We illustrate the  lower bounding process for a coupled Gaussian network $\mbN=(\prod_{i=1}^m \mX_i, p(\by|\bx), \prod_{j=1}^n \mY_j)$, where for all feasible $(i,j)$, the channel from $X_i$ to $Y_j$ has  SNR $\gamma_{ij}$ (incorporating the transmitted
signal power and the corresponding channel gain).
We first decompose the coupled network into decoupled MACs and BCs. For illustration purpose, we only focus on a pair of decoupled MAC and BC that share a common channel $\mX_i\to\mY_j$. Let  $\mbN_{i\to D_i}=(\mX_i, p(\by_{D_i}|x_i), \prod_{k\in D_i} \mY_k)$  be the decoupled BC with transmit signal $X_i$ and  $\mbN_{S_j\to j}=(\prod_{k\in S_j} \mX_k, p(y_j|\bx_{S_j}), \mY_j)$ be the decoupled MAC with receive signal $Y_j$. Without loss of generality, we assume that $D_i\subset\{1,\ldots,n\}$, $S_j\subset\{1,\ldots,m\}$ such that $i\in S_j$, $j\in D_i$, $|S_j|\geq 2$  and $|D_i|\geq 2$.

As in \textbf{Step II}, we first construct lower bounding models for the decoupled BC $\mbN_{i\to D_i}$ based on the rate and power allocation strategy as described in Sec.~\ref{sec:Eqv:BClower}. Depending on the channel quality, the messages carried by  $X_i$  may contain components that are not intended for decoding at the decoupled MAC  $\mbN_{S_j\to j}$.  The
 component(s) of $X_i$ that can't be decoded by  $\mbN_{S_j\to j}$ will
therefore behave as interference during the decoding process. We
denote the power of the interfering component from $X_i$ at $Y_j$ by $\Gamma_{ij}$, and
the exact value can be obtained from the lower bounding model of $\mbN_{i\to D_i}$.
 We  have $\Gamma_{ij}=0$ if all messages contained in $X_i$ are intended for
successful decoding, and $\Gamma_{ij}=\gamma_{ij}$ if nothing is to be
decoded.
After careful examination of  all  input
signals $\bx_{S_j}$, the total power of interfering
components contained in $Y_j$ is given by
\begin{align}\label{eqn:Eqv:P-I}
P^I_j = \sum_{k\in S_j}  \Gamma_{kj}.
\end{align}
 We can now construct the lower bounding model for the  decoupled MAC  $\mbN_{S_j\to j}$ in \textbf{Step III} based on the effective SNR, which is defined as
 \begin{align}\label{eqn:Eqv:Rmac-n}
\hat{\gamma}_{kj}= \frac{\gamma_{kj}-\Gamma_{kj}}{1+P^I_j}, \ \forall k\in S_j.
 \end{align}
For $k\in S_j$, let $l_k$ indicate the order of successive decoding in constructing the MAC lower bounding model: $l_i=1$ means $X_i$ is the first to be decoded. When decoding the message conveyed by  signal $X_i$,  the amount of interference introduced by signals other than $X_i$ is
\begin{align}\label{eqn:Eqv:P-In}  
P^I_{i,j} =  \sum_{\substack{k\in S_j \\ l_k < l_i}} \Gamma_{kj}  + \sum_{\substack{k\in S_j \\ l_k < l_i}} \gamma_{kj}.
\end{align}
We call $P^I_{i,j}$ the ``extrinsic interference'' of $X_i$ suffered during the decoding process of $Y_j$.

After {Step III}, we have obtained by \eqref{eqn:Eqv:P-In} the
extrinsic interference power $P^I_{i,k}$ caused by input signals
other than $X_i$ in the decoupled MAC with output signal $Y_k$, for all $k\in D_i$.
As suggested in \textbf{Step IV}, we should now update the lower bounding models for decoupled BC $\mbN_{i\to D_i}$ constructed in Step II by taking these extrinsic interference into account. To facilitate straightforward comparison with the lower bounding models developed in Sec.~\ref{sec:Eqv:BClower}, without loss of generality\footnote{This assumption always holds as we can set $\gamma_{ik}=0$ for all $k\not\in D_i$, and change labels when necessary.}, we assume $D_i=\{1,\ldots,n\}$ with
 $\gamma_{i1}\leq \ldots \leq \gamma_{in}$.

The rate constraint $R_{2^n-2^{l-1}}$ defined by
\eqref{eqn:Eqv:BC-Ri} in Sec.~\ref{sec:Eqv:BClower}, which corresponds to the multicast rate
  to a subset of receivers associated with $\{Y_k:
k=l,l+1,\ldots,n\}$, should be adjusted by taking into account the
extrinsic interference power $\{P^I_{i,k}: k=l,l+1,\ldots,n\}$. The
corresponding new rate constraint, denoted by $R'_{2^n-2^{l-1}}$,
is therefore defined as
\begin{align}\label{eqn:Eqv:BC-Ri2}
R'_{2^n-2^{l-1}} {=} \min_{k\in\{l,l+1,\ldots,n\}}
\frac{1}{2}\log\left(1 {+} \frac{ \gamma_{ik} \beta_l }{1{+} P^I_{i,k}
{+}\gamma_{ik}\sum_{t=l+1}^n \beta_t}\right).
\end{align}
The sum-rate constraint should be adjusted accordingly, i.e.,
\begin{align}
R'_0  = \sum_{l=1}^n R'_{2^n-2^{l-1}}.
\end{align}
Note that the minimum operation in \eqref{eqn:Eqv:BC-Ri2} comes
from the fact that given $\gamma_{il}\leq \ldots \leq \gamma_{in}$,
we can't guarantee
\begin{align}
\frac{\gamma_{il}}{1+ P^I_{i,l}} \leq \ldots \leq \frac{\gamma_{in}}{1+ P^I_{i,n}}
\end{align}
due to the effect of the extrinsic interference caused by
decoupled MACs. Here we simply keep the structure of the original
lower bounding model unchanged without claiming its optimality. 

Although we only demonstrate the two-step update via a Gaussian example, this approach is also feasible for non-Gaussian additive noise.
For point-to-point channel, Gaussian noise is the most pessimistic given the same SNR constraint. For decoding at  MACs, Gaussian noise is also the most pessimistic since we can always treat all the received signals as if they were sent from a single transmitter via superposition encoding with power allocation.

\begin{remark}
In our approach the effect of interference due to non-decoded messages has been incorporated in the two-step update process where successive interference cancellation decoding is used at all receiving nodes. Alternatively, one may treat the $m\times n$ coupled network together
and design a feasible coding approach with better interference management. For example, in \cite{Kannan14} an inner bound for the capacity region of a coupled  bidirectional network with ergodic symmetric channel fading coefficients has been characterized based on the ergodic interference alignment~\cite{Nazer09}. Further more, its inner bound is characterized only by ``local'' rate constraints associated with each node, which makes it feasible to be served as building blocks for large networks.
\end{remark}

\section{Illustrative Examples}\label{sec:coupled:illust}

 In this section we illustrate our capacity bounding models by several coupled noisy networks and compare the capacity inner and outer bounds obtained based on our bounding models with some benchmarks. Given a coupled noisy network $\mbN$, we first apply the channel decoupling method proposed in Sec.~\ref{sec:coupled:decoup} to decompose the coupled network
into decoupled  MACs and BCs. We then construct an upper bounding model $\mbC_{u,i}$ following the procedure described in Sec.~\ref{sec:Eqv:Upper} and a lower bounding model $\mbC_{l,i}$  as in Sec.~\ref{sec:Eqv:Lower}.
  As there are more than one upper (resp. lower) bounding model for
every MAC/BC, each of such combinations will result in a
  noiseless upper bounding model $\mbC_{u,i}$ (resp.
lower bounding model $\mbC_{l,i}$), whose capacity region serves as a capacity outer (resp. inner) bound for the original
noisy network $\mbN$.
We then
 take the intersection of all the capacity outer bounds, one for each
$\mbC_{u,i}$, to obtain the final (and tighter) outer bound,
\begin{align}\label{equ:upper:common}
  \mE(\mbN) \subseteq  \bigcap_i \mE(\mbC_{u,i}),
\end{align}
where $\mE(\cdot)$ denotes the capacity region of the corresponding network.
For the inner bound, we  compute the achievable rate region for
each lower bounding model $\mbC_{l,i}$ and
then take the convex hull of all the achievable rate regions to
create the final (and tighter) inner bound (with abuse of notation),
\begin{align}\label{equ:lower:union}
 \bigcup_i \mE(\mbC_{l,i}) \subseteq  \mE(\mbN).
\end{align}

\subsection{The Smallest Coupled Network: the Relay Channel}

\begin{figure}[t]
 \centering
 \psfrag{s1}[][]{$\mS'\ \ $}\psfrag{1d}[][]{$\ \ \mD'$}
\psfrag{s}[][]{$\mS$}\psfrag{d}[][]{$\mD$}\psfrag{r}[][]{$\mR$}
\psfrag{sd}[][]{$\gamma_{sd}$}\psfrag{sr}[][]{$\gamma_{sr}$}\psfrag{rd}[][]{$\gamma_{rd}$}
\psfrag{Rsd}[][]{$\max\{R_{l1},R_{d1}\}$}\psfrag{Rl2}[][]{$R_{l2}$}\psfrag{Rd2}[][]{$R_{d2}$}
\psfrag{ls}[][]{$R_{s}$}\psfrag{ds}[][]{$R_{d}$}
\psfrag{l1}[][]{$R_0$}\psfrag{l2}[][]{$R_1$}\psfrag{l3}[][]{$R_3$}
\psfrag{l4}[][]{$R_4$}\psfrag{l5}[][]{$R_5$}\psfrag{l6}[][]{$R_6$}
\psfrag{d1}[][]{$R'_0$}\psfrag{d2}[][]{$R'_2$}\psfrag{d3}[][]{$R'_3$}
\psfrag{d4}[][]{$R'_4$}\psfrag{d5}[][]{$R'_5$}\psfrag{d6}[][]{$R'_6$}
\psfrag{eq}{}
  \includegraphics[width=8.9cm]{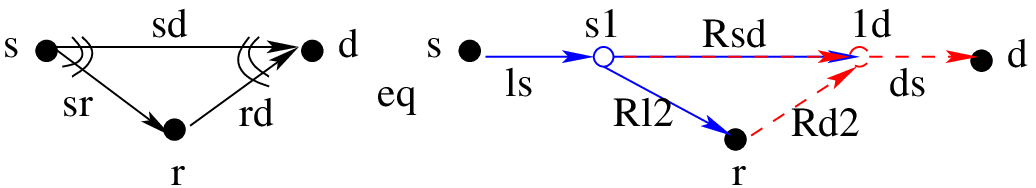}\\
  (a) Gaussian relay channel \hspace{4mm} (b) upper bounding model\\
  \includegraphics[width=8.6cm]{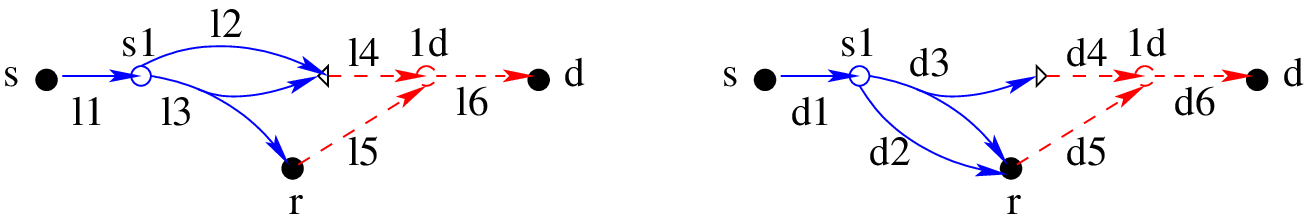}\\
  (c) for $\gamma_{sd}{>}\gamma_{sr}$ ---lower bounding models--- (d) for $\gamma_{sd}{<} \gamma_{sr}$
    \caption{The three-node Gaussian relay channel (a), its upper bounding model (b), and lower bounding models for the scenario when the
    source-destination link has stronger (c) or weaker (d) channel gain compared to the source-relay link.}
    \label{fig:NetEquvRelayLow}
\end{figure}

We first look at the smallest (in size) coupled network --- the
classical 3-node Gaussian relay channel --- as illustrated in
 Fig.~\ref{fig:NetEquvRelayLow}(a), which can be modelled as
 \begin{align*}
 Y & = \sqrt{\gamma_{sd}} X + \sqrt{\gamma_{rd}} X_r + Z,\\
 Y_r & = \sqrt{\gamma_{sr}} X + Z_r,
 \end{align*}
 where $\gamma_{sd}$, $\gamma_{sr}$, and $\gamma_{rd}$ are effective link SNRs, $Z$ and $Z_r$ are independent
  Gaussian noise with zero mean and unit variance, and $X$ and $X_r$ are transmitting signals subject to unit
  average power constraint.   The lower bounding model has two
types: we use the topology shown in
Fig.~\ref{fig:NetEquvRelayLow}(c) if the source-destination link
is better than the source-relay link, otherwise
Fig.~\ref{fig:NetEquvRelayLow}(d) is chosen.

\begin{figure}[t]
 \centering \includegraphics[width=8.6cm]{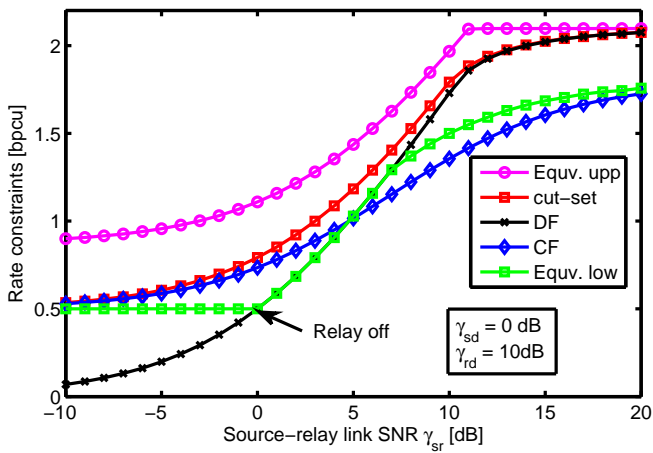}
    \caption{Upper and lower bounds on the capacity of a Gaussian relay channel
    with source-destination link SNR $\gamma_{sd}=0$dB and relay-destination link
    SNR $\gamma_{rd}=10$dB. Without resorting to source-relay cooperation,
    our equivalence lower bound demonstrates the general behavior of both
    CF (strong/weak $\gamma_{sr}$) and DF (medium $\gamma_{sr}$) while
    remaining a single bounding technique. }
    \label{fig:Neq:relay:rateg}
\end{figure}

The upper bounding model $(R_d, R_{d1}, R_{d2})$ for the decoupled
MAC can be constructed following the technique developed in
Sec.~\ref{sec:noncoup:MAC:upper} with link SNRs $\gamma_{sd}$ and
$\gamma_{rd}$. The upper bounding model $(R_s, R_{l1}, R_{l2})$
for the decoupled BC can be constructed following the technique
developed in Sec.~\ref{sec:Eqv:BC:upper} but with adjusted link
SNRs $\gamma_{sd}/\alpha$ and $\gamma_{sr}$. The optimal noise
partition parameter $\alpha$ is uniquely determined by the convex
optimization problem defined in \eqref{eqn:BC:NoisePart}, whose
solution for this special case can be solved analytically,\footnote{Following the same process as in Appendix~\ref{Proof:decouple:gauss} but focusing only on equations defined by \eqref{eqn:BC:Lagran}.}
\begin{align}
\alpha=\frac{\sqrt{\gamma_{sd}(1+\gamma_{rd})}}{\sqrt{\gamma_{rd}(1+\gamma_{sr}+\gamma_{sd})}+\sqrt{\gamma_{sd}(1+\gamma_{rd})}}.
\end{align}
 The connection between the auxiliary nodes $\mS'$ and $\mD'$
has two parallel rate constraints---$R_{l1}$ imposed by the
decoupled BC and $R_{d1}$ by the MAC---which results in the rate
constraint $\max\{R_{l1},R_{d1}\}$ as shown in
Fig.~\ref{fig:NetEquvRelayLow}(b).

In Fig.~\ref{fig:Neq:relay:rateg} we evaluate the  upper and lower
bounds obtained from our bounding models with respect to varying
source-relay link SNR $\gamma_{sr}$, and  compare them to three
benchmarks developed in~\cite{Cover79}: the cut-set upper bound,
the decode-and-forward (DF) lower bound, and the
compress-and-forward (CF) lower bound. Without resorting to
source-relay cooperation, our equivalence lower bound outperforms
CF and only suffers a loss of 0.3 bits per channel use compared to
DF when the source-relay link is strong.  When the source-relay
link is weak, it approaches the capacity and outperforms DF by
turning off the ``less capable'' relay node (when $\gamma_{sr}\leq
\gamma_{sd}$). Note that our equivalence lower bound has the
general behavior of both CF (strong/weak $\gamma_{sr}$) and DF
(medium $\gamma_{sr}$). Our equivalence upper bound, which assumes perfect
source-relay cooperation and relay-destination joint decoding, has
a gap of around 0.4 bits from the cut-set upper bound when
$\gamma_{sr}$ is small and the gap vanishes when $\gamma_{sr}$
becomes large.

\subsection{Multiple Unicast over a Layered Noisy Network}\label{sec:coupled:unicastM}

\begin{figure}[t]
\psfrag{w1}[][]{$W_{1}$}\psfrag{w2}[][]{$W_{2}$}\psfrag{wn}[][]{$W_{n}$}
\psfrag{v1}[][]{$\hat{W}_{1}$} \psfrag{v2}[][]{$\hat{W}_{2}$}\psfrag{vn}[][]{$\hat{W}_{n}$}
\psfrag{w3}[][]{$W_{n-1}$}\psfrag{v3}[][]{$\hat{W}_{n-1}$}\psfrag{s31}[][]{$R_{n-1,1}$}\psfrag{s32}[][]{$R_{n-1,2}$}
\psfrag{r1}[][]{$R_{}$}\psfrag{r2}[][]{$R_{s}$}\psfrag{r3}[][]{$R_{b}$}
\psfrag{r4}[][]{$R_{m}$}\psfrag{r8}[][]{$R_c$}
\psfrag{s11}[][]{$R_{1,1}$}\psfrag{s12}[][]{$R_{1,2}$}
\psfrag{s21}[][]{$R_{2,1}$}\psfrag{s22}[][]{$R_{2,2}$}
\psfrag{sn1}[][]{$R_{n,1}$}\psfrag{sn2}[][]{$R_{n,2}$}
 \centering \includegraphics[width=8cm]{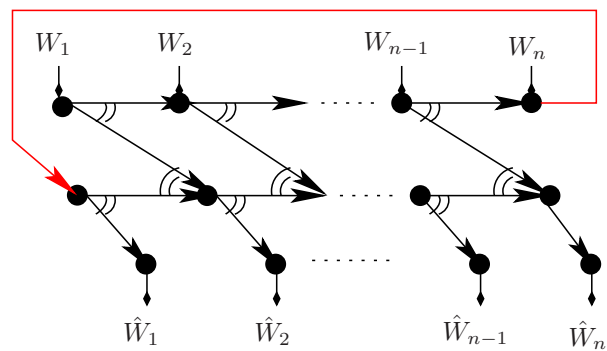}\\
 (a) multiple unicast over a coupled noisy network \\
  \includegraphics[width=8.6cm]{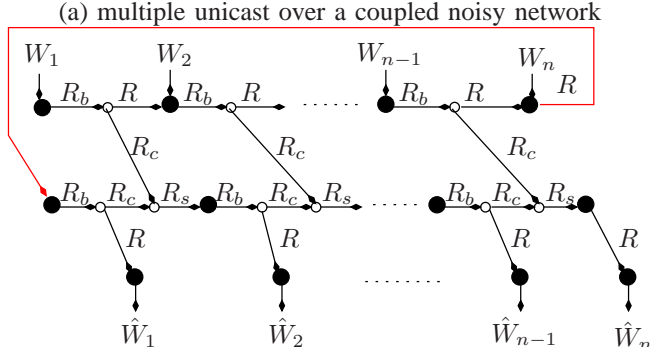}\\
  (b) upper bounding model with $R_c  = \max\{R_b, R'\}.$\\
   \includegraphics[width=8.6cm]{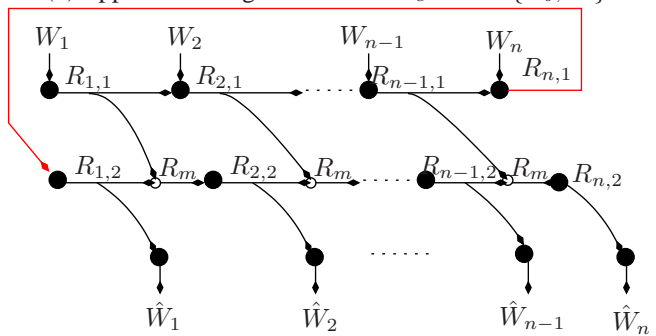}\\
  (c) lower bounding model
    \caption{Multiple unicast transmission over a coupled noisy network (a) where  the channel in red color
    is the bottleneck. Assuming all channels are Gaussian with identical link SNR $\gamma$,
    the upper bounding model (b) has only point-to-point bit pipes  and the lower bounding model (c)
    also contains point-to-points bit pipes (hyper-arcs) due to the broadcast transmission.}
    \label{fig:NetEquv_BCs}
\end{figure}

In Fig.~\ref{fig:NetEquv_BCs}(a) we focus on a layered multiple-unicast
noisy network where $n$ source-destination pairs are assisted by
$n$ intermediate  relaying nodes. It is a modified version
of~\cite[Fig.~3]{Effros2}\footnote{The network
of~\cite[Fig.~3]{Effros2} is obtained from the multiple-unicast
network in~\cite[Fig.~1]{unicastM} which consists of
point-to-point bit pipes only.} by replacing the orthogonal
transmissions to the relaying nodes with MACs to formulate a
coupled network. Assuming all transmission channels are identical
and all unicast rates are the same, the channel in red color is
the bottleneck. If all channels are Gaussian with link SNR
$\gamma$, we can construct a upper bounding model as shown in
Fig.~\ref{fig:NetEquv_BCs}(b) by first performing channel
decoupling as described in Sec.~\ref{sec:coupled:decoup}, and then substituting the upper bounding models developed in Sec.~\ref{sec:Eqv:BC:upper}
to replace BCs at source nodes by $\mbC_{u, BC, a}=(R_b,
R, R_b)$, BCs at intermediate relaying nodes by $\mbC_{u,
BC, b}=(R_b, R_b, R)$, and MACs by $\mbC_{u, MAC,
new}(\alpha)=(R_s,R',R')$, where
\begin{align}
R & = \frac{1}{2}\log\left(1+\gamma\right), \\
R_b &= \frac{1}{2}\log\left(1+3\gamma\right), \\
R' & = \frac{1}{2}\log\left(1+\frac{\gamma}{(1{-}\alpha)/2}\right),  \\
R_s &= \frac{1}{2}\log\left(1+\frac{4\gamma+1{-}\alpha}{\alpha}\right).
\end{align}
Note that although the value of $R_s$ and $R_c$ in
Fig.~\ref{fig:NetEquv_BCs}(b) may change by varying
$\alpha\in[0,1]$, the red-color link of capacity $R$ is always the
bottleneck, and therefore the symmetric-rate unicast capacity is
upper bounded by $R/n$.  This is actually the capacity as all the
source nodes can successfully transmit $B$ packets each over
$(B\cdot n +1)$ transmission blocks, which leads to a rate
\[\frac{B\cdot R}{B\cdot n +1} = \frac{R}{n+ 1/B} \to \frac{R}{n}, \mbox{ when } B\to\infty.\]
We illustrate the transmission scheme in
Table~\ref{tab:BCs:scheme} by an example with $n=4$.

\begin{table}[t]
\centering \caption{Transmission scheme with $n=4$ where source
node $\mS_i$ unicasts messages $W_i^b$, $b{=}1,\ldots,B$, to
destination node $\mD_i$ over $(n\cdot B{+}1)$ transmission blocks.
$W_i^0{=}\emptyset$ by default indicates no transmission.
 } \label{tab:BCs:scheme}
\begin{tabular}{|c|c@{$\mid$}c|c@{$\mid$}c|c@{$\mid$}c|c@{$\mid$}c|}
\hline
 $t= \backslash$ TX  & $\mS_1$ & $\mS_2$ & $\mS_3$ & $\mS_4$ & $\mR_1$    & $\mR_2$   & $\mR_3$   & $\mR_4$ \\
 \hline
  $4(b{-}1) {+} 1$       & $W_1^b$ & $W_2^b$ & $W_3^b$ & $W_4^b$ &$W_1^{b-1}$ &$W_2^{b-1}$&$W_3^{b-1}$&$W_4^{b-1}$\\
 \hline
  $4(b{-}1) {+} 2$       &$\emptyset$&$W_1^b$ &$W_2^b$ & $W_3^b$ & $W_4^b$ & $\emptyset$ & $\emptyset$& $\emptyset$\\
 \hline
  $4(b{-}1) {+} 3$    &$\emptyset$&$\emptyset$&$W_1^b$ &$W_2^b$ & $W_3^b$ & $W_4^b$ & $\emptyset$ & $\emptyset$\\
 \hline
  $4(b{-}1) {+} 4$    &$\emptyset$&$\emptyset$&$\emptyset$&$W_1^b$ &$W_2^b$ & $W_3^b$ & $W_4^b$ & $\emptyset$ \\
 \hline
\end{tabular}
\end{table}

The lower bounding model in in Fig.~\ref{fig:NetEquv_BCs}(c)
contains point-to-points bit pipes (hyper-arcs)  of rate $R_{i,k}
\leq  R$, for $i=1,\ldots,n, \ k{=}1,2$. By forcing successful
decoding at MACs, we will have the following constraints
\begin{align}
R_{i,1} + R_{i,2} \leq  R_m \triangleq \frac{1}{2}\log(1+2\gamma) , \forall i=1,\ldots,n-1. \label{eqn:BCs:Rsum}
\end{align}
Therefore we can conclude from the lower bounding model  that a
rate $R_l$ is  achievable if for all $i=1,\ldots,n,$
\begin{align}
i R_l & \leq  R_{i,1},  \label{eqn:BCs:Sr}\\
(n-i+1) R_l & \leq  R_{i,2}. \label{eqn:BCs:Rd}
\end{align}
After applying the MAC constraint \eqref{eqn:BCs:Rsum}, we have
\begin{align}
R_l &= \min\left\{\frac{R}{n}, \frac{R_m}{n+1}\right\} \\
    &= \left\{\begin{array}{cl}
            \frac{R}{n}, & \text{if } \gamma<\frac{1+\sqrt{5}}{2} \text{ or } n\geq\frac{\log(1+\gamma)}{\log(1+\gamma/(1+\gamma))},\\
            \frac{R_m}{n+1}, & \text{otherwise.}
            \end{array}
       \right.
\end{align}
Therefore the lower bounding models can provide a rate equals the capacity $R/n$ when the the SNR $\gamma$ is small or when the number of nodes is large.


\subsection{Multiple Multicast over a Wireless Relay Network}

\begin{figure}[t]
\psfrag{s1}[][]{$\mS_1$} \psfrag{s2}[][]{$\mS_2$}
\psfrag{d1}[][]{$\mD_1$}\psfrag{d2}[][]{$\mD_2$} \psfrag{dn}[][]{$\mD_n$}
\psfrag{x1}[][]{$X_1$} \psfrag{x2}[][]{$X_2$} \psfrag{xs}[][]{$X_s$}\psfrag{ys}[][]{$Y_s$}
\psfrag{y1}[][]{$Y_1$} \psfrag{y2}[][]{$Y_2$} \psfrag{yn}[][]{$Y_n$}
\psfrag{w1}[][]{$W_1$} \psfrag{w2}[][]{$W_2$}
\psfrag{w12}[][]{$\hat{W}_1\hat{W}_2$} \psfrag{w22}[][]{$\bar{W}_1\bar{W}_2$}\psfrag{wn2}[][]{$\tilde{W}_1\tilde{W}_2$}
\psfrag{a1}[][]{$\gamma_{11}$} \psfrag{a2}[][]{$\gamma_{12}$}\psfrag{an}[][]{$\gamma_{1n}$}
\psfrag{b1}[][]{$\gamma_{21}$} \psfrag{b2}[][]{$\gamma_{22}$}\psfrag{bn}[][]{$\gamma_{2n}$}
 \centering \includegraphics[width=8.6cm]{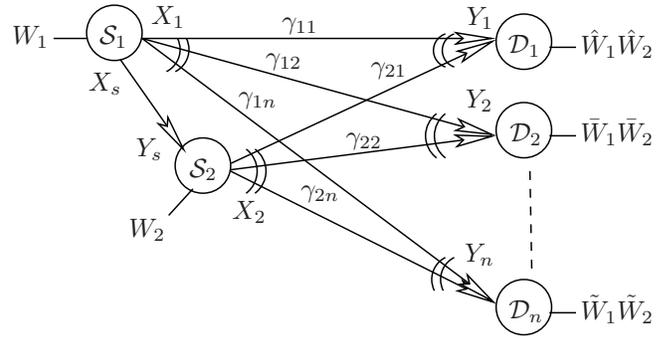}
    \caption{Two source nodes $\mS_1$ and $\mS_2$ multicast
    $W_1$ and $W_2$ respectively to all destinations $\{\mD_i: i=1,... ,n\}$ through Gaussian channels,
    where $\gamma_{ji}$ denotes the  SNR of the link from $\mS_j$ to $\mD_i$. The link $(X_s\to Y_s)$
    from $\mS_1$ to $\mS_2$ is a $Q$-ary symmetric channel orthogonal to all the other Gaussian channels.}
    \label{fig:S1S2Dn}
\end{figure}

\begin{figure}[t]
\psfrag{s1}[][]{$\mS_1$} \psfrag{s2}[][]{$\mS_2$}
\psfrag{d1}[][]{$\mD_1$}\psfrag{d2}[][]{$\mD_2$} \psfrag{d3}[][]{$\mD_3$}
\psfrag{w1}[][]{$W_1$} \psfrag{w2}[][]{$W_2$}
\psfrag{r11}[][]{$R'_{1}$}  \psfrag{r13}[][]{$R'_{3}$} \psfrag{r17}[][]{$R'_{7}$}
\psfrag{r24}[][]{$R_{4}$} \psfrag{r26}[][]{$R_{6}$}\psfrag{r27}[][]{$R_{7}$}
\psfrag{c1}[][]{$C_{12}$}
\psfrag{m1}[][]{$R_{11}$}\psfrag{m2}[][]{$R_{21}$}
\psfrag{m3}[][]{$R_{12}$}\psfrag{m4}[][]{$R_{22}$}
\psfrag{m5}[][]{$R_{13}$}\psfrag{m6}[][]{$R_{23}$}
 \centering \includegraphics[width=8.6cm]{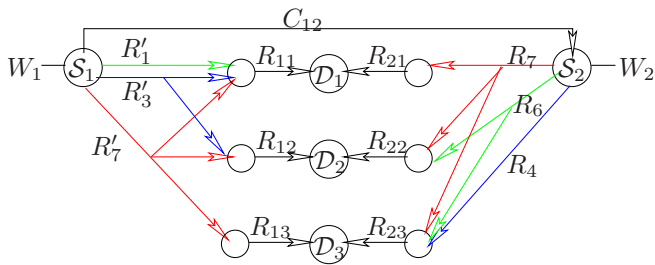}
    \caption{Lower bounding model for the scenario with $n=3$ destinations where $C_{12}$ is the channel capacity of the $Q$-ary symmetric channel $(X_s\to Y_s)$. $R'_3$ and $R'_7$ are the rates of hyper-arcs from $\mS_1$ to $\{\mD_1\mD_2\}$ and $\{\mD_1\mD_2\mD_3\}$, respectively, and $R_6$ denotes the multicast rate from $\mS_2$ to $\{\mD_2\mD_3\}$. Rate constraints $R_{sd}$, $s=1,2$ and $d=1,2,3$, come from the lower bounding models of MACs.}
    \label{fig:S1S2Dn:low}
\end{figure}

We now illustrate the construction of
bounding models for a multiple multicast relay network shown in
Fig.~\ref{fig:S1S2Dn}, where two source nodes $\mS_1$ and
$\mS_2$
multicast information $W_1$ at rate $R_1$ and $W_2$ at rate $R_2$,
respectively, to all destinations $\{\mD_i: i=1,... ,n\}$ over
Gaussian channels. The effective SNRs of these Gaussian channels are
\begin{align}
\gamma_{1k} = P - \frac{k}{n}\Delta_P, \nonumber\\
\gamma_{2k} = P + \frac{k}{n}\Delta_P, \label{eqn:S1S2Dn:P}
\end{align}
where $P>\Delta_P>0$ are parameters such that
\begin{align}
\gamma_{2n}>\cdots>\gamma_{21}>\gamma_{11}>\cdots>\gamma_{1n}.
\end{align}
 The transmission of $\mS_1$ is aided by $\mS_2$ via a $Q$-ary symmetric channel $(X_s\to Y_s)$ such that for all $m,k\in\{0,1,\ldots,Q-1\}$
 \begin{align}
 Pr(Y_s=m|X_s=k)=\left\{\begin{array}{ll}
   1-\xi, &  \mbox{ if } m=k ,\\
   \frac{\xi}{Q-1} & \mbox{ if } m\neq k.
   \end{array}\right.
 \end{align}
We can first decompose the original network into a point-to-point
channel, two decoupled BCs originating from $\mS_1$ and $\mS_2$, and $n$ decoupled MACs ending at each destination node,
and then construct
upper and lower bounding models as described in Sec~\ref{sec:Eqv:coup}. An illustration of the resulting lower bounding model for $n=3$ is presented in Fig.~\ref{fig:S1S2Dn:low}, where the rate constraints of poin-to-point(s) bit pipes are determined following the process  in Sec.~\ref{sec:Eqv:Lower}. From Fig.~\ref{fig:S1S2Dn:low} we can see that multicast of $W_2$ can only be achieved via the hyper-arc of rate $R_7$, while multicast of $W_1$ can be achieved either via the hyper-arc of rate $R'_7$, or via the collaboration with $\mS_2$ at rates $\min\{R'_3, R_4\}$ and/or $\min\{R'_1, R_6\}$. The collaboration from $\mS_2$ is subject to the rate constraint $C_{12}$ which is the capacity of the $Q$-array symmetric channel from $\mS_1$ to $\mS_2$.

\begin{figure}[t]
\centering \includegraphics[width=8.6cm]{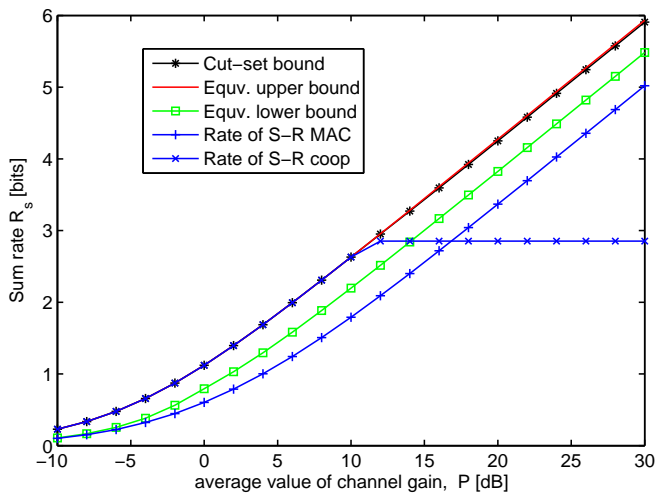}
 \caption{\label{Eqv:fig:PN10} Bounds on the sum rate for the scenario with $10$ destinations, $\delta_P/P=-3$dB, and the $Q$-ary symmetric channel $(X_s\to Y_s)$ with $Q=8$ and $\xi=0.1$.}
\end{figure}

 The bounds on sum rate obtained from
upper and lower bounding networks with respect to varying channel quality have been illustrated in
Fig.~\ref{Eqv:fig:PN10} for a scenario with $n=10$ destinations. The $Q$-ary symmetric channel $(X_s\to Y_s)$ has parameters $Q=8$ and $\xi=0.1$, which results in a capacity of $C_{12}=2.85$ bits per channel use [bpcu]. We also plot three benchmarks as
 references: the rate achieved by transmitting identical signals from $\mS_1$ and $\mS_2$ (denoted by ``S1-S2 coop''), the rate
 achieved by transmitting independent signals from $\mS_1$ and $\mS_2$ (denoted by ``S1-S2 MAC''), and the cut-set upper bound following the method developed in~\cite{Cover79}\footnote{In~\cite{Cover79} the cut-set bound is obtained by starting from multi-letter expressions and then combining
 various inequalities, average power constraints, and some ``properly'' chosen auxiliary random variables. Here ``properly'' is to highlight the fact that it is a kind of art to decide when and
 where to introduce auxiliary random variables to quantify the
 possible correlation among transmitting signals, since only
 proper choice leads to nice upper bounds. See~\cite{NC_relay,
 NC_X_relay,JSAC11} for extensions of the method developed in~\cite{Cover79}.}.
Our upper bound obtained from noiseless bounding networks is
very good\footnote{In the sense that the gap from the
cut-set bound is negligible.} and it even approaches the capacity (meeting the lower bound provided by $\mS_1$-$\mS_2$ cooperation) in low to medium SNR regions. Our lower bounding models
 discard the possibility of source cooperation and therefore
 suffers some performance degradation (less than 0.4 bits\footnote{As $P$ and $n$ increase, the gap from cut-set bound converges to a constant that depends only on $\Delta_P/P$.} from the
 capacity). In high SNR region, it outperforms the two benchmarks since our lower bounding models can make use of the
 overhead messages to increase the multicast rate of $W_1$: extra bits of $W_1$ can be transmitted via the collaboration with $\mS_2$ at rate
 \begin{align}
 \Delta_{R1} = \min\{R'_3, R_4\} + \min\{R'_1, R_6\},
 \end{align}
 if such operation is permitted by the link from $\mS_1$ to $\mS_2$, i.e., when $R_4+R_6<C_{12}$.

\section{Summary}\label{sec:Eqv:conc}

In this work we have presented capacity upper and lower bounding
models for wireless networks, where the upper bounding models
consist of only point-to-point bit pipes while the lower bounding
models also contains point-to-points bit pipes (hyper-arcs). We
have extended the bounding models for two-user MACs/BCs to
many-user scenarios and established a constant additive gap between upper and
lower bounding models. For networks with coupled links, we have
proposed a channel decoupling method to decompose
coupled networks into  decoupled MACs and BCs, and
proposed strategies to construct step-by-step upper and
lower bounding models for the originally coupled networks.  We have demonstrated by examples that the gap between the resulting
upper and lower bounds is usually not large, and the upper/lower
bound can approach capacity in some setups. The proposed methods
for constructing upper and lower bounding models, effective and
computationally efficient, can be easily extended to large
networks with complexity grows linearly with the number of nodes. They therefore, combined with methods calculating the
capacity of noiseless networks, provide  additional powerful tools
for characterizing the capacity region of general wireless
networks.

\appendices

\section{Optimal Noise Partitioning for Gaussian MACs}\label{app:Noise}

Following the Lagrangian method~\cite{Convex}, the optimal noise partitioning for the optimization problem \eqref{eqn:MAC:NoisePart} can be obtained by taking partial derivatives of its Lagrangian
\begin{align}
L = \sum_i \log(1+\gamma_i/\alpha_i) + \mu^{-1}(\sum_i \alpha_i - (1-\alpha)), \mbox{ } \mu>0,
\end{align}
with respect to $\alpha_i$, $i=1,\ldots,m$, and setting them to zero.  Denoting $\alpha_i^*$ the optimal noise power for $\alpha_i$, we have
\begin{align}
 (\alpha_i^*)^2 + \gamma_i \alpha_i^* -\gamma_i\mu =0,
\end{align}
which leads to (omitting the negative root as $\alpha_i^*>0$)
\begin{align}
\alpha_i^*  = \frac{\sqrt{\gamma_i(\gamma_i+4\mu)}-\gamma_i}{2}.
\end{align}
The exact value of $\mu$ is determined by the condition $\sum_{i=1}^m \alpha_i =1-\alpha$, which yields \eqref{eqn:MAC:noise:mu}.

\section{Proof of Lemma~\ref{lemma:MAC:mu}}\label{Proof:lemma:MAC:mu}

 The upper bound is obtained by
contradiction. Assuming $\mu > \frac{1-\alpha}{m} +
\frac{(1-\alpha)^2}{m^2}\frac{1}{\min_i \gamma_i}$, the LHS of
\eqref{eqn:MAC:noise:mu} is evaluated as follows
\begin{align}
\mbox{LHS } & {>}  \frac{1}{2}\sum_{j=1}^m  \Bigg(\sqrt{\left(\gamma_j {+} \frac{1{-}\alpha}{m/2}\right)^2 {+} \frac{(1{-}\alpha)^2}{m^2} \left(\frac{\gamma_j}{\min_i \gamma_i}{-}1\right)}{-}\gamma_j\Bigg)\nonumber\\
 & \geq  \frac{1}{2}\sum_{j=1}^m  \left(\sqrt{\left(\gamma_j + \frac{1-\alpha}{m/2}\right)^2 }-\gamma_j\right) \\
 & = 1-\alpha, \nonumber
\end{align}
which contradicts to the equality constraint stated in
\eqref{eqn:MAC:noise:mu}. Therefore we have
\begin{align}
\mu \leq \frac{1-\alpha}{m} + \frac{(1-\alpha)^2}{m^2}\frac{1}{\min_i \gamma_i},
\end{align}
where the equality holds if and only if $\gamma_j {=} \min_i
\gamma_i$ for all $j{=}1,\ldots,m$, i.e., when
$\gamma_1=\ldots=\gamma_m$.

The lower bound is obtained as follows. From
\eqref{eqn:MAC:noise:mu} we have
\begin{align}
\left(2(1-\alpha)+ \sum_i \gamma_i\right)^2 & = (\sum_i \sqrt{\gamma_i(\gamma_i+4\mu)} )^2 \\
                      & \leq \left(\sum_i \gamma_i\right) (4m\mu+\sum_i \gamma_i ) \label{eqn:app:CSi}\\
                      & = 4m\mu \sum_i\gamma_i +  (\sum_i\gamma_i )^2,
\end{align}
where the inequality in \eqref{eqn:app:CSi} is due to Cauchy-Schwarz inequality with
equality holds if and only if $\gamma_1=\ldots=\gamma_m$. By
expanding the LHS and removing common items at both sides, we can
easily obtain the lower bound.

\section{Proof of Lemma~\ref{lemma:MAC:Rs}}\label{proof:lemma:MAC:Rs}

From \eqref{eqn:MAC:new:sum} it is straightforward to observe that
$R_s (\alpha)$ is a monotonously decreasing function with respect
to $\alpha$, with its minimum
\begin{align}
R_s(\alpha{<}1)> R_s(\alpha{=}1)=R_{MAC}\triangleq \frac{1}{2}\log\left(1 {+}  \left(\sum_{i=1}^m \sqrt{\gamma_i}\right)^2\right).
\end{align}
Therefore we only need to prove $\sum_{i=1}^m R_i(\alpha)>R_{MAC}$.

From \eqref{eqn:MACRs:alpha} and \eqref{eqn:MAC:noise:mu}it is
easy to see that  $R_i (\alpha)$  is a monotonously increasing function with respect to
$\alpha\in[0,1)$, which leads to
\begin{align}
\sum_{i=1}^m R_i(\alpha>0) > \sum_{i=1}^m R_i(\alpha=0).
\end{align}

On the other hand, we have
\begin{align}
\sum_{i=1}^m R_i(\alpha=0) & = \sum_{i=1}^m \frac{1}{2}\log\left(1 + \frac{\gamma_i}{\alpha_i}\right) \\
                & > \frac{1}{2} \log\left(1 + \sum_{i=1}^m  \frac{\gamma_i}{\alpha_i} \right) \\
                & \geq \frac{1}{2}\log\left(1 + \min_{x_i>0;\sum x_i=1} \sum_{i=1}^m \frac{\gamma_i}{x_i} \right) \\
                & = \frac{1}{2}\log\left(1 + \left (\sum_{i=1}^m \sqrt{\gamma_i}\right)^2 \right) \label{eqn:apx:opt:x}\\
                & = R_{MAC},
\end{align}
where \eqref{eqn:apx:opt:x} holds by the optimal solution
$x^*_i=\sqrt{\gamma_i}/(\sum_j \sqrt{\gamma_j})$. Hence we have proved the lemma.

\section{Noise Partition for Gaussian Channel Decoupling}\label{Proof:decouple:gauss}

Let
\begin{align}\label{eqn:BC:Lagr}
L= \sum_{i=1}^n \log\left(1+ \sum_{j=1}^{m} \frac{\gamma_{i,j}}{\alpha_{i,j}}\right) + \sum_{j=1}^m \lambda_j\left(\sum_{i=1}^n \alpha_{i,j}-1\right),
\end{align}
be the Lagrangian, by taking partial derivative of $L$ w.r.t.
$\alpha_{i,j}$ and setting them to zero, we  get
\begin{align}\label{eqn:BC:Lagran}
\frac{\gamma_{i,j}}{\alpha^2_{i,j}} =  \lambda_j\left(1 + \sum_{k=1}^{m}\frac{\gamma_{i,k}}{\alpha_{i,k}}\right).
\end{align}
By introducing auxiliary variables
\begin{align}
\mu_i = 1 + \sum_{j=1}^{m}\frac{\gamma_{i,j}}{\alpha_{i,j}}, \ \forall i\in\{1,\ldots,n\}, \label{eqn:BC:Lagr:mu}
\end{align}
we can derive from \eqref{eqn:BC:Lagran} the following equations
\begin{align}
\alpha_{i,j} & =  \frac{\sqrt{\gamma_{i,j}}}{\sqrt{\lambda_j}\sqrt{\mu_i}},\label{eqn:BC:Lagr:alpha}\\
\frac{\gamma_{i,j}}{\alpha_{i,j}} & = \sqrt{\gamma_{i,j}}\sqrt{\lambda_j}\sqrt{\mu_i}, \label{eqn:BC:Lagr:SNR}\\
\sqrt{\lambda_j} & = \sum_{i=1}^n  \frac{\sqrt{\gamma_{i,j}}}{\sqrt{\mu_i}},   \label{eqn:BC:Lagr:lambda}\\
\mu_i & = 1 + \sqrt{\mu_i}\sum_{j=1}^{m}  \sqrt{\gamma_{i,j}}\sqrt{\lambda_j},  \label{eqn:BC:Lagr:mu2}\\
\sqrt{\mu_i}  & = \frac{1}{2}\left( \sqrt{\left(\sum_{j=1}^m \sqrt{\gamma_{i,j}}\sqrt{\lambda_j}\right)^2+4} + \sum_{j=1}^m \sqrt{\gamma_{i,j}}\sqrt{\lambda_j} \right), \label{eqn:BC:Lagr:musq}
\end{align}
where \eqref{eqn:BC:Lagr:lambda} comes from
\eqref{eqn:BC:Lagr:alpha} and the fact that $\sum_{i=1}^n
\alpha_{i,j}=1$,
\eqref{eqn:BC:Lagr:mu2} is obtained by substituting \eqref{eqn:BC:Lagr:SNR} into \eqref{eqn:BC:Lagr:mu}, and \eqref{eqn:BC:Lagr:musq} is the unique feasible solution to \eqref{eqn:BC:Lagr:mu2}.
Therefore the equivalent SNRs $\frac{\gamma_{i,j}}{\alpha_{i,j}}$
for decoupled BCs are uniquely determined by
\eqref{eqn:BC:Lagr:SNR} where the optimal value of $\mu_i$ and
$\lambda_j$ can be easily obtained by iterating
\eqref{eqn:BC:Lagr:lambda} and \eqref{eqn:BC:Lagr:musq}. The
convergence to the global optimum is guaranteed by observing the
fact that $\lambda_j$ is a monotonically decreasing function of
$\{\mu_i: \forall i\}$ via \eqref{eqn:BC:Lagr:lambda} and $\mu_i$
is a monotonically increasing function of $\{\lambda_j: \forall
j\}$, via \eqref{eqn:BC:Lagr:musq}.

\end{document}